\documentclass[aps,prb,preprint,showpacs]{revtex4}
\usepackage{natbib}
\bibpunct{(}{)}{,}{a}{}{,}
\usepackage{epsfig}
\begin{document}
\title{Exact ground states for the four electron problem in a two-dimensional
finite Hubbard square system.}
\author{Endre~Kov\'acs and Zsolt~Gul\'acsi}  
\affiliation{Department of Theoretical Physics, University of Debrecen, 
H-4010 Debrecen, Hungary}
\date{\today}
\begin{abstract}
We present exact explicit analytical results describing the exact ground state
of four electrons in a two dimensional square Hubbard cluster containing 16 
sites taken with periodic boundary conditions. The presented procedure, which 
works for arbitrary even particle number and lattice sites, is based on
explicitly given symmetry adapted base vectors constructed in 
${\bf r}$ space. The Hamiltonian acting on these states generates a closed 
system of 85 linear equations providing by its minimum eigenvalue the exact 
ground state of the system. The presented results, described with the aim to 
generate further creative developments, not only show how the ground state 
can be exactly obtained and what kind of contributions enter in its 
construction, but emphasize further characteristics of the spectrum. On this 
line i) possible explications are found regarding why weak coupling 
expansions often provide a good approximation for the Hubbard model at  
intermediate couplings, or ii) explicitly given low lying energy states of the
kinetic energy, avoiding double occupancy, suggest new roots for pairing
mechanism attracting decrease in the kinetic energy, as emphasized by
kinetic energy driven superconductivity theories. 
\end{abstract}
\maketitle

\section{Introduction}

The growing interest in systems of highly correlated electrons has proposed 
many theoretical descriptions where the dominant effects are treated in the 
frame of the Hubbard model \citep{E2D1}. Since the exact solution of the
model is known only in one dimension \citep{E2D2}, and the majority of driving
forces in the field, as superconductivity in cuprates, are two dimensional 
(2D) problems, a great variety of approximate treatments have been proposed
\citep{E2D3} in order to accommodate a suitable theoretical framework. 
Despite several years of intensive studies, it is apparent that the necessary
theoretical skills and tools to deal with this problem are still in fact 
relatively poor \citep{E2D4}.

While most of the ongoing effort relating the Hubbard model is being 
concentrated on the general case of large electron density, more than ten years
ago it has been recognized that the less analysed low-density limit not only
that retain main aspects related to the model behaviour \citep{E2D5,E2D5a}, 
but can provide key knowledge which could drive further at least 
nonperturbative developments \citep{E2D6}, firstly because is more hope to find
non-approximated solutions holding by their nature essential information. 
The history of this background has started with the exact solution of the 
two-electron ground state problem on an arbitrary large torus 
\citep{E2D7,E2D7a},
solution of the two electron problem in 2D \citep{E2D8}, and continued with the
three electron problem extended nonperturbatively in the low density limit
\citep{E2D6}. The numerical treatment of the four electron problem on four 
sites follows by the test of the Bardeen-Cooper-Schrieffer and 
resonant-valence-bond wave functions as approximated ground states of the 
Hubbard model \citep{E2D5}, test of the unrestricted Hartree-Fock solution
\citep{E2D9}, the calculation of Huckel-Hubbard correlation diagrams 
\citep{E2D10}, and deduction of absorption spectra of TCNQ particles
\citep{E2D5a}. The concurrently made numerical developments  on $4 \times 4$ 
clusters around half filling \citep{E2D11,E2D12} corroborated by 
group-theoretical studies of the $4 \times 4$ square system \citep{E2D13}
also lead to the development of the numerical description of the $4$ 
particle problem for energy level statistics \citep{E2D14,E2D15}. In the last 
years the study of symmetry properties for Hubbard clusters in general
has continued \citep{E2D16}, extensions to Hubbard-Holstein model for clusters 
holding 4 electrons has been given \citep{E2D17}, and for spinless fermion case
numerical characterization of the 4 particle problem goes up to clusters of 
$20 \times 20$ extension \citep{E2D18}. Despite the invested efforts, exact 
analytical results holding essential information are not known at the moment 
in the problem. 

The last period has provided strong new motivations for the further development
of the study of the low concentration limit of the Hubbard model. The reason 
for this is that experimentally one started to encounter in condensed matter
context rapidly increasing situations containing small number of particles
confined in a system or device, as for example in the case of
quantum dots \citep{i1}, quantum well structures \citep{i2}, mesoscopic 
systems \citep{d5},  experimental entanglement \citep{i5} etc. Between the 
studied experiments, several are directly connected to the Hubbard model, as 
in the case of charge transfer complexes \citep{End1}, 
quantum dots \citep{End2},
or mesoscopic grains \citep{End3}. Furthermore, it has been observed that 
several measurements on strongly correlated systems are acceptable reproduced
on small Hubbard clusters, as in the case of the X-ray absorption for 
nickelates studied on 4-site cluster \citep{End4}. The same can be stated for
the $4 \times 4$ cluster case used in describing manganites \citep{End5,End6},
photoemission intensities \citep{End7}, or thermodynamic properties provided
by processes in the vicinity of the Fermi surface \citep{End8}.

As can be observed, the accumulated knowledge to the present date 
regarding the extreme
low density limit described by few particles present in the system, excepting 
the one and two particle cases, means only symmetry properties, and numerical
results (often approximated), which usually conceal explicit properties which 
could generate
creative advancement at the level of the theoretical description and deep
understanding. This is a regrettable and unfortunate situation, since as 
is known from the accumulated experience in solving exactly  
many-body systems \citep{End9}, or as
several times has been accentuately stressed \citep{E2D4,E2D6}, key aspects of
the unapproximated descriptions are often hidden in the few particle cases.

On the presented background, with the aim to fill up at least partially
this gap for the two
dimensional Hubbard case, and being driven by the intention to provide 
essential information for the subjects mentioned above, we present the exact 
ground state of four interacting electrons in a $4 \times 4$ Hubbard cluster.
Our main purpose is not to hide potential essential characteristics behind
numerical results, but to present explicit expressions, basic resulting
properties, and visible
characteristics, which we strongly hope, are able to polarize the creative 
thinking, and advancement in the field. 

Our results are based on an ${\bf r}$ space description, which in our opinion
is unjustly overshaded in the last period in treating such problems, in 
contradiction with its deep ability to bring to light essential 
characteristics. We construct first a base vector set starting from symmetry
properties, and then show how a closed system of linear equations containing 
85 components can be constructed characterizing the ground state manifold and
providing by its secular equation, through its minimum eigenvalue, the ground 
state wave function and ground state energy. After this step the properties
of the ground state are analyzed. The concretely described situation is an 
$L\times L =N=16$ square lattice with periodic boundary conditions in both 
directions, and $N_p=4$ particles. But the method itself works for arbitrary 
even number of particles and arbitrary even $L$. The explicitly presented
base vectors and equations providing the ground state, not only show how the
ground state can be constructed and what kind of components build it up, but
provides an insight into other properties of the spectrum as well. Especially 
the characteristics regarding the kinetic energy eigenstates should be 
mentioned on this line, important for understanding aspects related to
perturbative expansions, or kinetic energy driven superconductivity.

The remaining part of the paper is structurated as follows. Section II. 
presents the Hamiltonian, the deduction procedure and the ground state wave 
functions, Section III. describes physical properties of the deduced
eigenstates, Sect. IV. presents the conclusions of the paper, while the 
Appendices A - B presenting mathematical details, close the presentation.

\section{Hamiltonian and ground state wave functions.}
\subsection{Presentation of the Hamiltonian}

The Hamiltonian we use has the form of a standard Hubbard Hamiltonian
\begin{eqnarray}
\hat H = - t \sum_{<i,j>,\sigma} (\hat c^{\dagger}_{i,
\sigma} \hat c_{j,\sigma} + H.c.) 
+ U \sum_{i} \hat n_{i,\uparrow} \hat n_{i,\downarrow} ,
\label{ez1}
\end{eqnarray}
where $\hat c^{\dagger}_{i,\sigma}$ creates an electron at site $i$ with spin
$\sigma$, $t$ is the nearest-neighbour hopping
amplitude, $U$ is the on-site Coulomb repulsion, and $<i,j>$ represents 
nearest-neighbour sites, taken into account in the sum over sites only once.
During our study we consider an $L \times L$ cluster, $N=L^2$, with periodic 
boundary conditions in both directions. All the presented results will relate
$\hat H/t$, the only one microscopic parameter of the problem being $u=U/t$.
The case $N=16$ will be presented in details, although the method is applicable
for arbitrary even $N$. Since $N_p=4$, the band filling for the presented case
is $0.125$. We further mention that we consider during this paper $u > 0$.

\subsection{The construction of the base wave vectors}

\subsubsection{The basic wave vector elements}

In order to describe the Hilbert space region containing the ground state
for the $N_p=4$ particle problem, we use an ${\bf r}$-space representation 
for the wave vectors. In order to characterize this, first the lattice sites 
of the considered system are numbered starting
from the down-left corner, the numbering being given along the lower lattice 
sites line, then going upward with the notation, as shown in Fig.1.a. 
In the considered system (taking into account that the ground state 
is a singlet state), three type of
particle configurations may occur as depicted in Fig.1.b,c,d.

 1) We can have
two double occupancies at sites $i$ and $j$ represented by two dots at sites
$i$ and $j$ (see Fig.1.b). 2) We may have a double occupancy at site $i$ and 
two electrons with opposite spins at sites $j$ and $k$. In this case, the 
double occupancy at site $i$ it is denoted as before by a dot at $i$, while 
the two electrons with opposite spins placed on the different sites ($j$, and 
$k$) are depicted by a dotted line connecting the sites $j$ and $k$ 
(see Fig.1.c). Finally, we may have four single 
occupancies placed on different sites. From these, the electrons placed at $i$
and $j$ have spin $\sigma$, while the electrons at sites $k$ and $l$ have spin
$-\sigma$. This situation will be represented by two continuous lines 
connecting the sites $i$ and $j$, and $k$ and $l$, respectively (see Fig.1.d).
The notations $i,j,k,l$ are representing the numbering of the sites as 
specified in Fig.1.a.

The mathematical expressions connected to the states represented in Fig.1.b,c,d
are given as follows. The two double occupancies present in Fig.1.b provide
the state
\begin{eqnarray}
|b\rangle = (\hat c^{\dagger}_{i,\uparrow} \hat c^{\dagger}_{i,\downarrow})
(\hat c^{\dagger}_{j,\uparrow} \hat c^{\dagger}_{j,\downarrow}) |0\rangle,
\label{ez2}
\end{eqnarray}
where $|0\rangle$ represents the bare vacuum with no fermions present. The one
double occupancy and the two electrons with opposite spins depicted in Fig.1.c
are described as
\begin{eqnarray}
|c\rangle = (\hat c^{\dagger}_{i,\uparrow} \hat c^{\dagger}_{i,\downarrow})
[(\hat c^{\dagger}_{j,\uparrow} \hat c^{\dagger}_{k,\downarrow}) +
(\hat c^{\dagger}_{k,\uparrow} \hat c^{\dagger}_{j,\downarrow})] |0\rangle
\: .
\label{ez3}
\end{eqnarray}
For the here presented situation we must have $j \ne k$ and $i \ne j, i \ne k$
respectively. Finally, in describing mathematically the four single occupancies
presented in Fig.1.d, we have
\begin{eqnarray}
|d\rangle = [(\hat c^{\dagger}_{j,\uparrow} \hat c^{\dagger}_{i,\uparrow})
(\hat c^{\dagger}_{l,\downarrow} \hat c^{\dagger}_{k,\downarrow}) +
(\hat c^{\dagger}_{j,\downarrow} \hat c^{\dagger}_{i,\downarrow})
(\hat c^{\dagger}_{l,\uparrow} \hat c^{\dagger}_{k,\uparrow})] |0\rangle,
\label{ez4}
\end{eqnarray}
where for the mathematical clarity, $j > i$ and $l > k$ is required (all sites
being considered different).

\subsubsection{The translation of the basic elements}

In order to construct the base vectors,  the basic wave vector elements are 
translated by the operation $T$. The $T$ operator is considered a linear 
operator, hence the relation
\begin{eqnarray}
T(A + B) = T(A) + T(B) ,
\label{ez5}
\end{eqnarray}
holds, where $A$ and $B$ represent particle configurations as depicted in 
Fig.1.b,c,d.
Furthermore $T(A)$ means that the configuration of particles represented
by $A$ is translated to each site of the system (for $N=16$, the translation is
made 16 times), and the obtained contributions are all added (see Fig.2). 
The argument of the $T$ operator contains usually four terms since the
starting particle configuration is rotated by 180 degrees along the $x,y$, and
$z$ axes, and then the obtained contributions are added. Less than four terms
in the $T$ argument means that the mentioned rotations provide the same 
starting particle microconfiguration.

The translations are such effectuated that the relative inter-particle 
positions 
are all maintained, and arriving near the border of the system, the presence 
of the periodic boundary conditions are explicitly taken into account. The 
translations must be given at the level of the graphic plots, all the obtained
$16$ graphics must be added (see Fig.2), then each graphic must be transformed
in a mathematical form according to the rules presented in 
Eqs.(\ref{ez2}-\ref{ez4}). At this point we must underline, that when the 
mathematical expression is written for a given graph, a negative sign may 
emerge in some cases given by the restrictions prescribed for 
Eqs.(\ref{ez2}-\ref{ez4}).
For example, the last two terms in both rows of Fig.3 have mathematical
expressions with negative sign.
Similar reasons lead to sign changes obtained after rotations, reason which 
generates the sign changes observed in some $T(...)$ arguments of base vectors
presented in Figs.4-18, for example in the case of the second contributions of
$|10\rangle$, or $|11\rangle$, etc. Furthermore, in some cases (see for example
$|12\rangle$ or $|15\rangle$), the rotations provide the starting 
microconfiguration of particles, which leads to the decrease of the number
of terms present in the argument of the $T$ operation. The sign changes also 
lead in some cases to complete elimination of some possible particle 
microconfigurations.

\subsubsection{The base wave vectors}

Starting from the rules and operations described in details in Sec.II.B.1-2,
the 85 base wave vectors can be explicitly constructed as given in Fig.4-18.
These are denoted by $|n\rangle$, $n=1,2,...,85$, and are all orthogonal
vectors. As can be seen, the base wave vectors are obtained by subtracting
two type of contributions. The first part of these is obtained from a 
starting 
particle microconfiguration (presented as first terms in the argument of the 
first operator $T$) which is rotated by 180 degrees along the $x$, $y$ and 
$z$ axes, and translated by the operation $T$ after this step. The second 
(subtracted) part of the wave vectors is obtained from
the first part by a rotation of the elements by 90 degree along the $z$ axis.

\subsection{The ground state wave function}

First of all we are constructing a closed system of equations containing the 
base vectors described above. 

This system of 85 equations is obtained as
follows. i) We start from the most {\em condensed} and most {\em interacting} 
particle microconfiguration depicted by the base vector $|1\rangle$ 
(see Fig.4), which cannot be eliminated on physical grounds from the ground 
state. ii) By applying the Hamiltonian on the base vector $|1\rangle$
(see the first equation of Appendix A), we obtain as a result new base vectors
($|2\rangle$ and $|3\rangle$), which have the same symmetry properties
as $|1\rangle$. iii) Since the Hamiltonian does not change the symmetry
properties of the base vectors, one must further apply $\hat H$ on the each
resulting new base vector, up to the moment in which the obtained system of 
equations closes. This happens for the studied system at the 85th equation.
The obtained closed system of equations is explicitly presented in Appendix A. 
The minimum energy solution of the presented system of equations represents
the ground state of the system.

\section{Properties of the solutions}

\subsection{The check of the ground state energies}

First of all one can check that the system of equations presented in Appendix A
indeed provides the ground state of the problem. The obtained minimum 
eigenvalues provided by Appendix A are
\begin{eqnarray}
&&u = 0.0, \quad E = -12.0000000000000000,
\nonumber\\
&&u = 0.5, \quad E = -11.9126285145094126,
\nonumber\\
&&u = 1.0, \quad E = -11.8364690235642218,
\nonumber\\
&&u = 1.5, \quad E = -11.7695877912829268,
\nonumber\\
&&u = 2.0, \quad E = -11.7104580743242632,
\nonumber\\
&&u = 2.5, \quad E = -11.6578605150913841,
\nonumber\\
&&u = 3.0, \quad E = -11.6108110503797608,
\nonumber\\
&&u = 3.5, \quad E = -11.5685079573602838,
\nonumber\\
&&u = 4.0, \quad E = -11.5302924026297138 ,
\label{ez6}
\end{eqnarray}
which perfectly coincide with the numerically exact minimum eigenvalues 
obtained from exact numerical diagonalization in the 14 400 dimensional full
Hilbert space \citep{obs}. We note that the eigenvalues are given in $t$ units.

\subsection{The $U$ independent eigenstates}

The study of the solutions of the system of equations presented in
Eq.(\ref{a1}) leads to an extremely interesting observation,
namely that almost half (exactly 40\%) of the eigenstates are $U$ independent,
consequently are eigenstates of the non-interacting system as well. For
example, the $U$ independent $18$ eigenvectors corresponding to zero 
eigenvalue are presented in Appendix B. Besides these, one has also $16$
eigenfunctions corresponding to eigenvalues $\pm 4, \pm 8$, which are also
$U$ independent (see exemplification in Appendix B).
The presence of such eigenstates is important for following reasons.

First of all, several authors observed \citep{E2D12,vol} that weak coupling 
expansions often provide a good approximation for the Hubbard model at 
intermediate coupling. A possible explication for this is the emergence of a 
huge number of eigenstates in the spectrum of the Hubbard model with non-zero 
interaction, which are present in the non-interacting case as well in the 
spectrum of the model.

A second aspect which must be mentioned here is that the eigenstates related
to zero eigenvalue presented in Appendix B are in fact many-body eigenstates 
of the kinetic energy term with no double occupancy. Such states 
apparently totally avoid double occupancy at no cost of energy, consequently are important in the study of the pairing mechanism in the Hubbard model \citep{c1,c2,c3}, 
and as seen from the presented results, easily emerge in the spectrum.
 
At this step we underline that contrary to the practice used in the
theoretical studies up to this moment \citep{c4},
double occupancy avoiding eigenstates emerge not only at zero energy, but also
at much lower energy values, closely situated to the ground state energy. We
exemplify this statement by the eigenstate $|v_2\rangle$ presented in 
Eq.(\ref{b2}), corresponding to energy $-8$ (in $t$ units). These eigenstates 
being $U$ independent, their energy remain unchanged if the strength
of the on-site interaction is increased, while the ground-state energy 
increases if $U$ is increased. Consequently, starting from these states,
the contribution in the pairing mechanism of
the kinetic energy eigenstates not containing double occupancy could be much
more efficiently taken into account than considered up today. We note that
this problem has interconnections to the problem of the kinetic energy driven 
superconductivity as well, strongly debated in the literature in the last 
period \citep{s1,s2,s3,s4}. Indeed, experimentally is observed \citep{s5,s6}
that especially in $Bi$ based cuprates, a decrease in the kinetic energy is 
observed at the superconducting transition, which contradicts the usual
(for example BCS) pairing theories \citep{s1}. If the pairing process could be
described on the states of the type $|v_2\rangle$, exemplified in our knowledge
in exact terms for the first time in this paper, a such a decrease could be
much better understood.

\subsection{The $U$ dependent eigenstates}

The major part of the spectrum (60\%) is $U$ dependent. From the eigenstates 
entering in this category, there are  states whose eigenvalue is 
exactly $u$, for example 
$|w\rangle=|8\rangle - |9\rangle + |39\rangle + |40\rangle-|41\rangle$.
Such states are interesting for two reasons. First, a detection possibility of
such states would lead to a direct experimental measurement of the $U/t$ ratio.
Second, since all the contributions in $|w\rangle$ have double occupancy $d=1$,
this vector is also an eigenstate of the interaction term with eigenvalue $u$,
and in the same time, the eigenvector of the kinetic energy term with 
eigenvalue zero. Consequently, the study of the zero kinetic energy eigenvalue
states must be given with care, since such type of states, as exemplified
before, not necessarily are avoiding the double occupancy.

The remaining part of the deduced $U$ dependent states, are not eigenstates 
of the kinetic energy, and the interacting part of $\hat H$ separately,
but are eigenvectors only for the sum of both, e.g. the whole Hamiltonian.
As seen from Eq.(\ref{ez6}), the corresponding eigenvalues have a smooth 
(less than linear) $U$ dependence, and the ground states is obtained from this
category.

\subsection{The system generating the ground state}

In order to enhance further developments and enlighten further creative
thinking, we provide explicitly in Appendix A the equations generating the 
ground state, together with the explicit form of the base vectors providing 
the ground state (see Figs.4-18), hence describing the Hilbert space region
in which this is placed. We mention that if the base vectors are defined only 
through the operation $T$, without subtracting two $T$ components as shown in
Figs.4-18, a system containing 176 equations arises. This however can be cast 
in two block diagonal components, one of which provides the equations 
presented in Appendix A. 

\section{Summary and Conclusions}

Driven by the aim to provide explicit expressions generating creative 
developments, we present exact analytical results describing the ground 
state of the 
two dimensional Hubbard model taken on an $L \times L =N$ square lattice
at $N=16$ (and periodic boundary conditions in both directions), containing 
$N_p=4$ particles. The presented procedure allows the 
description of an arbitrary even $L$ and $N_p$, and is based on symmetry 
adapted base vectors constructed in ${\bf r}$ space. The Hamiltonian acting on
the described base vectors provides a closed system of linear equations (whose
number is 85 for $N=16$ and $N_p=4$), leading by its secular equation, 
through its minimum eigenvalue, to the ground state wave function and 
ground state energy of the system, deduced in a Hilbert subspace with almost 
three orders of magnitude smaller in dimensions than the full Hilbert space of 
the problem. The deduced explicit eigenstates characterize also other 
properties of the spectrum: i) The large number of eigenstates which remain
eigenstates of the non-interacting Hamiltonian as well shows why weak coupling 
expansions often provide a good approximation for the Hubbard model at  
intermediate coupling. ii) Zero energy eigenstates of the kinetic energy term
($\hat H_{kin}$)
which are eigenstates of the Hamiltonian as well show how energy increasing
double occupancies can be avoided providing a possible support for the kinetic
energy driven superconductivity. iii) Low energy eigenstates of the kinetic
energy term which completely avoid double occupancy emphasize potentially new
pairing possibilities in the low energy part of the spectrum in the context of
the kinetic energy driven superconductivity. iv) Zero energy 
eigenstates of the kinetic energy term corresponding to double occupancy 
one underline
that the zero energy $\hat H_{kin}$ eigenstates must be 
handled with care, since as exemplified, these states not necessarily represent
double occupancy avoiding states.

Similarly obtained solutions for higher $L$ or $N_p$, corroborated with the 
study of the emerging changes in the system of equations describing the 
ground state, remain a challenge for future developments.
  
\acknowledgments

This work was supported by the Hungarian Scientific Research 
Fund through contract OTKA-T-037212. The numerical calculations have been 
done at the Supercomputing Lab. of the Faculty of Natural Sciences, 
Univ. of Debrecen, supported by OTKA-M-041537. Z. G. kindly acknowledge
test numerical results obtained from exact numerical diagonalization on the 
full Hilbert space provided by Sandro Sorella and Arnd Hubsch.

\appendix

\section{The linear system of equations containing the ground state.}
\def\theequation{{\thesection}\arabic{equation}}
This Appendix presents the 85 linear equations describing the ground state 
of the studied system. 

\begin{eqnarray}
&&\hat H |1\rangle= 2 u |1\rangle + |2\rangle - |3\rangle ,
\nonumber\\
&&\hat H |2\rangle=|1\rangle + u |2\rangle + 8|4\rangle - |6\rangle 
- |7\rangle + |10\rangle + |11\rangle + 4 |12\rangle ,
\nonumber\\
&&\hat H |3\rangle= -8 |1\rangle + u |3\rangle - 8|5\rangle + |6\rangle 
+ |7\rangle -2 |8\rangle -2 |9\rangle - |10\rangle - |11\rangle -2|13\rangle
-2 |14\rangle -4 |15\rangle ,
\nonumber\\
&&\hat H |4\rangle=|2\rangle + 2u |4\rangle - |17\rangle ,
\nonumber\\
&&\hat H |5\rangle=-|3\rangle + |23\rangle ,
\nonumber\\
&&\hat H |6\rangle= -2 |2\rangle + |3\rangle + u |6\rangle - |16\rangle + 
2 |17\rangle - 2 |18\rangle - |19\rangle + |20\rangle + 2 |21\rangle - 
|23\rangle + |24\rangle 
\nonumber\\
&&\hspace*{1cm}+ |25\rangle + |26\rangle +|29\rangle + |31\rangle,
\nonumber\\
&&\hat H |7\rangle= -2 |2\rangle + |3\rangle + u |7\rangle - |16\rangle + 
2 |17\rangle + |20\rangle +2 |22\rangle - |23\rangle + |25\rangle + 
|26\rangle + |28\rangle + |30\rangle ,
\nonumber\\
&&\hat H |8\rangle= - |3\rangle + u |8\rangle - |16\rangle - 
|21\rangle - |22\rangle + |23\rangle - |27\rangle - |33\rangle - 
|34\rangle ,
\nonumber\\
&&\hat H |9\rangle= - |3\rangle + u |9\rangle - |16\rangle - 
|19\rangle - |20\rangle + |23\rangle - |24\rangle - |27\rangle - 
|32\rangle ,
\nonumber\\
&&\hat H |10\rangle= 2 |2\rangle - |3\rangle + |16\rangle - 2|17\rangle - 
2 |18\rangle - |19\rangle + |23\rangle + |24\rangle - |25\rangle - 
|26\rangle - |28\rangle 
\nonumber\\
&&\hspace*{1cm}- |31\rangle - |32\rangle -2 |33\rangle,
\nonumber\\
&&\hat H |11\rangle= 2 |2\rangle - |3\rangle + |16\rangle - 2|17\rangle + 
|23\rangle - |25\rangle - |26\rangle - |29\rangle - |30\rangle - 
|32\rangle - 2 |34\rangle , 
\nonumber\\
&&\hat H |12\rangle = |2\rangle - |25\rangle ,
\nonumber\\
&&\hat H |13\rangle= - |3\rangle - |16\rangle + |19\rangle + 
|23\rangle + |24\rangle - |27\rangle - |28\rangle - |29\rangle ,
\nonumber\\ 
&&\hat H |14\rangle= - |3\rangle - |16\rangle + |23\rangle - 
|27\rangle - |30\rangle - |31\rangle ,
\nonumber\\ 
&&\hat H |15\rangle= - |3\rangle - |26\rangle - |27\rangle , 
\nonumber\\ 
&&\hat H |16\rangle= - |6\rangle - |7\rangle - 2 |8\rangle - 2|9\rangle + 
|10\rangle + |11\rangle - 2 |13\rangle - 2 |14\rangle + u |16\rangle - 
8 |36\rangle 
\nonumber\\
&&\hspace*{1cm}+ 4 |46 \rangle - 4 |54\rangle , 
\nonumber\\
&&\hat H |17\rangle= -8 |4\rangle + |6\rangle + |7\rangle - |10\rangle - 
|11\rangle + u |17\rangle + 4 |36\rangle -2 |40\rangle - 2 |41\rangle - 
4 |44\rangle 
\nonumber\\
&&\hspace*{1cm}-4 |47\rangle + 2 |63\rangle + 2|64\rangle ,
\nonumber\\
&&\hat H |18\rangle= - |6\rangle - |10\rangle + u |18\rangle + |42\rangle - 
|56\rangle + |58\rangle ,
\nonumber\\
&&\hat H |19\rangle= - |6\rangle - 2 |9\rangle - |10\rangle + 2 |13\rangle + 
u |19\rangle - 2 |35\rangle - 4|39\rangle - 2 |42\rangle - 2 |55\rangle + 
|57\rangle + |58\rangle , 
\nonumber\\
&&\hat H |20\rangle=  |6\rangle + |7\rangle - 2 |9\rangle + u |20\rangle - 
2 |35\rangle - 2 |38\rangle - 4 |40\rangle - 2 |42\rangle - |43\rangle - 
|50\rangle - |51\rangle 
\nonumber\\
&&\hspace*{1cm} + |57\rangle + 2 |61\rangle + 2 |65\rangle ,
\nonumber\\
&&\hat H |21\rangle= |6\rangle - |8\rangle + u |21\rangle - 2|41\rangle + 
|42\rangle - |43\rangle - |50\rangle - |56\rangle - |59\rangle , 
\nonumber\\
&&\hat H |22\rangle= |7\rangle - |8\rangle + u |22\rangle - 2|38\rangle - 
2 |41\rangle - |51\rangle - |59\rangle ,  
\nonumber\\
&&\hat H |23\rangle= 8 |5\rangle - |6\rangle - |7\rangle + 2|8\rangle + 
2 |9\rangle + |10\rangle + |11\rangle + 2 |13\rangle + 2 |14\rangle +
2 |37\rangle
\nonumber\\
&&\hspace*{1cm} + 2 |38\rangle + 2 |43\rangle + 8 |44\rangle - 4 |45\rangle 
- 4 |46\rangle ,
\nonumber\\
&&\hat H |24\rangle= |6\rangle - 2|9\rangle + |10\rangle + 2|13\rangle - 
2 |35\rangle - 4 |39\rangle - 2 |55\rangle - 2 |56\rangle + |57\rangle -
|58\rangle, 
\nonumber\\
&&\hat H |25\rangle= |6\rangle + |7\rangle - |10\rangle - |11\rangle - 
8 |12\rangle + 4 |45\rangle + 4 |48\rangle + 4 |54\rangle -2 |59\rangle -
2 |60\rangle 
\nonumber\\
&&\hspace*{1cm} + 2 |61\rangle + 2 |62\rangle , 
\nonumber\\
&&\hat H |26\rangle= |6\rangle + |7\rangle - |10\rangle - |11\rangle - 
4 |15\rangle - 8 |47\rangle + 4 |48\rangle - |49\rangle - |50\rangle -
|51\rangle - |52 \rangle ,  
\nonumber\\
&&\hat H |27\rangle= -2|8\rangle -2 |9\rangle -2 |13\rangle -2 |14\rangle - 
4 |15\rangle + 4 |45\rangle - |49\rangle - |50\rangle - |51\rangle -
|52\rangle 
\nonumber\\
&&\hspace*{1cm} - 8 |53\rangle - 4 |54\rangle , 
\nonumber\\
&&\hat H |28\rangle= |7\rangle - |10\rangle -2 |13\rangle + 2 |35\rangle - 
2 |38\rangle + 2 |42\rangle - |43\rangle - |51\rangle - |52\rangle -
|57\rangle 
\nonumber\\
&&\hspace*{1cm} + 2 |62\rangle + 4 |63\rangle - 2 |65\rangle , 
\nonumber\\
&&\hat H |29\rangle= |6\rangle - |11\rangle -2 |13\rangle + 2 |35\rangle - 
2 |37\rangle - |43\rangle - |49\rangle - |50\rangle + 2 |56\rangle -
|57\rangle 
\nonumber\\
&&\hspace*{1cm} + 2 |62\rangle + 4 |63\rangle - 2 |65\rangle , 
\nonumber\\
&&\hat H |30\rangle= |7\rangle - |11\rangle -2 |14\rangle - 2 |37\rangle - 
2 |38\rangle - |49\rangle - |51\rangle - 2 |60\rangle + 4 |64\rangle ,
\nonumber\\
&&\hat H |31\rangle= |6\rangle - |10\rangle -2 |14\rangle - 2 |43\rangle - 
|50\rangle - |52\rangle - 2 |60\rangle + 4 |64\rangle ,
\nonumber\\
&&\hat H |32\rangle= - 2 |9\rangle - |10\rangle - |11\rangle - 2 |35\rangle - 
2 |37\rangle - 4 |40\rangle - |43\rangle - |49\rangle - |52\rangle -
2 |56\rangle 
\nonumber\\
&&\hspace*{1cm} + |57\rangle + 2 |61\rangle + 2 |65\rangle , 
\nonumber\\
&&\hat H |33\rangle= -|8\rangle - |10\rangle -2 |41\rangle - |42\rangle - 
|43\rangle - |52\rangle + |56\rangle - |59\rangle ,
\nonumber\\
&&\hat H |34\rangle= -|8\rangle - |11\rangle -2 |37\rangle -2 |41\rangle - 
|49\rangle - |59\rangle ,
\nonumber\\
&&\hat H |35\rangle= -|19\rangle - |20\rangle - |24\rangle + |28\rangle + 
|29\rangle - |32\rangle + |74\rangle - |75\rangle ,
\nonumber\\
&&\hat H |36\rangle= -|16\rangle + |17\rangle + 2u |36\rangle + |66\rangle , 
\nonumber\\
&&\hat H |37\rangle= |23\rangle - |29\rangle - |30\rangle - |32\rangle - 
2|34\rangle + 2 |67\rangle - |70\rangle + |71\rangle ,
\nonumber\\
&&\hat H |38\rangle= -|20\rangle -2 |22\rangle + |23\rangle - |28\rangle - 
|30\rangle + 2 |67\rangle - |70\rangle + |71\rangle ,
\nonumber\\
&&\hat H |39\rangle= -|19\rangle - |24\rangle + u |39\rangle ,
\nonumber\\
&&\hat H |40\rangle= - |17\rangle - |20\rangle - |32\rangle + u |40\rangle - 
|66\rangle +  |67\rangle - |69\rangle ,
\nonumber\\
&&\hat H |41\rangle= - |17\rangle - |21\rangle - |22\rangle - |33\rangle - 
|34\rangle + u  |41\rangle - |66\rangle + |67\rangle - |69\rangle ,
\nonumber\\
&&\hat H |42\rangle= |18\rangle - |19\rangle - |20\rangle + |21\rangle + 
|28\rangle  - |33\rangle + u  |42\rangle + |75\rangle - |76\rangle ,
\nonumber\\
&&\hat H |43\rangle= - |20\rangle - 2 |21\rangle + 2 |23\rangle - |28\rangle - 
|29\rangle  - 2 |31\rangle - |32\rangle - 2 |33\rangle + 4  |67\rangle 
- 2 |70\rangle + 2 |71\rangle ,
\nonumber\\
&&\hat H |44\rangle= - |17\rangle + |23\rangle + |67\rangle ,
\nonumber\\
&&\hat H |45\rangle= - |23\rangle + |25\rangle + |27\rangle + |70\rangle ,
\nonumber\\
&&\hat H |46\rangle= |16\rangle - |23\rangle - |71\rangle ,
\nonumber\\
&&\hat H |47\rangle= - |17\rangle - |26\rangle - |69\rangle ,
\nonumber\\
&&\hat H |48\rangle= |25\rangle + |26\rangle - |72\rangle ,
\nonumber\\
&&\hat H |49\rangle= - |26\rangle - |27\rangle - |29\rangle - |30\rangle - 
|32\rangle  - 2 |34\rangle - 2 |68\rangle - 2 |69\rangle -  |70\rangle 
+ |72\rangle - |73\rangle ,
\nonumber\\
&&\hat H |50\rangle= - |20\rangle - 2 |21\rangle - |26\rangle - |27\rangle
- |29\rangle - |31\rangle  - 2 |68\rangle - 2 |69\rangle - |70\rangle 
+ |72\rangle 
\nonumber\\
&&\hspace*{1cm} - |73\rangle + |74\rangle - |75\rangle - 2 |76\rangle ,
\nonumber\\
&&\hat H |51\rangle= - |20\rangle - 2 |22\rangle - |26\rangle - |27\rangle
- |28\rangle - |30\rangle  - 2 |68\rangle - 2 |69\rangle - |70\rangle 
+ |72\rangle  - |73\rangle ,
\nonumber\\
&&\hat H |52\rangle= - |26\rangle - |27\rangle - |28\rangle - |31\rangle
- |32\rangle - 2 |33\rangle  - 2 |68\rangle - 2 |69\rangle - |70\rangle 
+ |72\rangle 
\nonumber\\
&&\hspace*{1cm} - |73\rangle - |74\rangle + |75\rangle + 2 |76\rangle ,
\nonumber\\
&&\hat H |53\rangle= - |27\rangle - |68\rangle ,
\nonumber\\
&&\hat H |54\rangle = - |16\rangle + |25\rangle - |27\rangle - |73\rangle , 
\nonumber\\
&&\hat H |55\rangle = - |19\rangle - |24\rangle + |74\rangle + |75\rangle , 
\nonumber\\
&&\hat H |56\rangle = - |18\rangle - |21\rangle - |24\rangle + |29\rangle
- |32\rangle + |33\rangle + |74\rangle + |76\rangle , 
\nonumber\\
&&\hat H |57\rangle = |19\rangle + |20\rangle + |24\rangle - |28\rangle
- |29\rangle + |32\rangle - |74\rangle - |75\rangle , 
\nonumber\\
&&\hat H |58\rangle = 2 |18\rangle + |19\rangle - |24\rangle - |74\rangle
+ |75\rangle + 2 |76\rangle , 
\nonumber\\
&&\hat H |59\rangle = - |21\rangle - |22\rangle - |25\rangle - |33\rangle
- |34\rangle - |70\rangle + |72\rangle + |73\rangle , 
\nonumber\\
&&\hat H |60\rangle = - |25\rangle - |30\rangle - |31\rangle - |70\rangle
+ |72\rangle + |73\rangle , 
\nonumber\\
&&\hat H |61\rangle = |20\rangle + |25\rangle + |32\rangle + |70\rangle
- |72\rangle - |73\rangle - |74\rangle - |75\rangle , 
\nonumber\\
&&\hat H |62\rangle = |25\rangle + |28\rangle + |29\rangle + |70\rangle
- |72\rangle - |73\rangle + |74\rangle + |75\rangle , 
\nonumber\\
&&\hat H |63\rangle = |17\rangle + |28\rangle + |29\rangle + |66\rangle
- |67\rangle + |69\rangle , 
\nonumber\\
&&\hat H |64\rangle = |17\rangle + |30\rangle + |31\rangle + |66\rangle
- |67\rangle + |69\rangle , 
\nonumber\\
&&\hat H |65\rangle = |20\rangle - |28\rangle - |29\rangle + |32\rangle ,
\nonumber\\
&&\hat H |66\rangle= 4 |36\rangle - 2 |40\rangle - 2 |41\rangle + 2 |63\rangle
+ 2 |64\rangle + u |66\rangle  + 4 |78\rangle - 4 |85\rangle ,
\nonumber\\
&&\hat H |67\rangle= 2 |37\rangle + 2 |38\rangle + 2 |40\rangle + 2 |41\rangle
+ 2 |43\rangle + 4 |44\rangle  - 2 |63\rangle - 2 |64\rangle + 16 |77\rangle 
\nonumber\\
&&\hspace*{1cm} - 4 |78\rangle + 4 |79\rangle ,
\nonumber\\
&&\hat H |68\rangle= - |49\rangle - |50\rangle - |51\rangle - |52\rangle
- 4 |53\rangle - 8 |80\rangle  + 4 |81\rangle , 
\nonumber\\
&&\hat H |69\rangle= - 2 |40\rangle - 2 |41\rangle - 4 |47\rangle - |49\rangle
- |50\rangle - |51\rangle  - |52\rangle + 2 |63\rangle + 2 |64\rangle 
- 4 |79\rangle 
\nonumber\\
&&\hspace*{1cm} - 8 |80\rangle - 4 |85\rangle ,
\nonumber\\
&&\hat H |70\rangle= - 2 |37\rangle - 2 |38\rangle - 2 |43\rangle + 4
|45\rangle - |49\rangle - |50\rangle  - |51\rangle - |52\rangle - 2 |59\rangle 
- 2 |60\rangle 
\nonumber\\
&&\hspace*{1cm} + 2 |61\rangle + 2 |62\rangle - 8 |79\rangle + 8 |82\rangle
- 4 |84\rangle ,
\nonumber\\
&&\hat H |71\rangle= 2 |37\rangle + 2 |38\rangle + 2 |43\rangle - 4 |46\rangle
- 8 |78\rangle + 4 |84\rangle  ,
\nonumber\\
&&\hat H |72\rangle= - 4 |48\rangle + |49\rangle + |50\rangle + |51\rangle
+ |52\rangle + 2 |59\rangle  + 2 |60\rangle - 2 |61\rangle - 2 |62\rangle 
\nonumber\\
&&\hspace*{1cm} - 8 |81\rangle - 8 |82\rangle ,
\nonumber\\
&&\hat H |73\rangle= - |49\rangle - |50\rangle - |51\rangle - |52\rangle
- 4 |54\rangle + 2 |59\rangle  + 2 |60\rangle - 2 |61\rangle - 2 |62\rangle 
\nonumber\\
&&\hspace*{1cm} + 4 |84\rangle - 8 |85\rangle ,
\nonumber\\
&&\hat H |74\rangle= 2 |35\rangle + |50\rangle - |52\rangle + 2 |55\rangle
+ 2 |56\rangle - |57\rangle  - |58\rangle - 2 |61\rangle +2 |62\rangle 
+ 4 |83\rangle , 
\nonumber\\
&&\hat H |75\rangle= 2 |35\rangle + 2 |42\rangle - |50\rangle + |52\rangle
+ 2 |55\rangle - |57\rangle  + |58\rangle - 2 |61\rangle + 2 |62\rangle 
+ 4 |83\rangle , 
\nonumber\\
&&\hat H |76\rangle= - |42\rangle - |50\rangle + |52\rangle + |56\rangle
+ |58\rangle ,
\nonumber\\
&&\hat H |77\rangle= |67\rangle ,
\nonumber\\
&&\hat H |78\rangle= |66\rangle - |67\rangle - |71\rangle ,
\nonumber\\
&&\hat H |79\rangle= |67\rangle - |69\rangle - |70\rangle ,
\nonumber\\
&&\hat H |80\rangle= - |68\rangle - |69\rangle , 
\nonumber\\
&&\hat H |81\rangle= |68\rangle - |72\rangle ,
\nonumber\\
&&\hat H |82\rangle= |70\rangle - |72\rangle ,
\nonumber\\
&&\hat H |83\rangle= |74\rangle + |75\rangle ,
\nonumber\\
&&\hat H |84\rangle= - |70\rangle + |71\rangle + |73\rangle ,
\nonumber\\
&&\hat H |85\rangle= - |66\rangle - |69\rangle - |73\rangle .
\label{a1}
\end{eqnarray}

\section{The $U$ independent eigenvectors corresponding to zero eigenvalue.}

The $U$ independent eigenvalues corresponding to zero eigenvalues are the 
following
\begin{eqnarray}
&&|\theta_1\rangle = |35\rangle + |57\rangle ,
\nonumber\\
&&|\theta_2\rangle = |55\rangle + |57\rangle - |65\rangle ,
\nonumber\\
&&|\theta_3\rangle = |37\rangle - |38\rangle + |51\rangle - |49\rangle,
\nonumber\\
&&|\theta_4\rangle = |60\rangle + |61\rangle - |63\rangle + |64\rangle -
|65\rangle + |83\rangle,
\nonumber\\
&&|\theta_5\rangle = |60\rangle + |62\rangle - |63\rangle + |64\rangle -
|83\rangle,
\nonumber\\
&&|\theta_6\rangle = |77\rangle + |78\rangle - |80\rangle - |81\rangle +
|82\rangle + |84\rangle + |85\rangle,
\nonumber\\
&&|\theta_7\rangle = - |77\rangle + |79\rangle - |80\rangle - |81\rangle +
|82\rangle,
\nonumber\\
&&|\theta_8\rangle = |10\rangle - |11\rangle + |49\rangle - |52\rangle +
|58\rangle,
\nonumber\\
&&|\theta_9\rangle = |13\rangle - |14\rangle + |63\rangle - |64\rangle +
|55\rangle - |83\rangle,
\nonumber\\
&&|\theta_{10}\rangle = - |5\rangle + |15\rangle + |44\rangle - |47\rangle -
|53\rangle - |77\rangle + |80\rangle,
\nonumber\\
&&|\theta_{11}\rangle = |14\rangle - |15\rangle + |46\rangle - |48\rangle -
|60\rangle + |84\rangle,
\nonumber\\
&&|\theta_{12}\rangle = |44\rangle + |45\rangle - |47\rangle - |48\rangle +
|53\rangle - 2 |77\rangle + |79\rangle + |81\rangle,
\nonumber\\
&&|\theta_{13}\rangle = |29\rangle - |28\rangle + |30\rangle - |31\rangle +
|75\rangle - |74\rangle + 2 |33\rangle - 2 |34\rangle - 2 |76\rangle,
\nonumber\\
&&|\theta_{14}\rangle = - |45\rangle + |46\rangle + |54\rangle + 2 |77\rangle 
- 2 |53\rangle - 2 |79\rangle + 2 |80\rangle + |84\rangle,
\nonumber\\
&&|\theta_{15}\rangle = |37\rangle - |44\rangle + |47\rangle - |49\rangle -
|77\rangle + |53\rangle + |80\rangle - |82\rangle - |84\rangle,
\nonumber\\
&&|\theta_{16}\rangle = |43\rangle - 2 |37\rangle + 2 |49\rangle - |50\rangle 
- |52\rangle,
\nonumber\\
&&|\theta_{17}\rangle = |10\rangle - 2 |12\rangle - |15\rangle -2 |44\rangle -
|45\rangle - |48\rangle - |52\rangle + |54\rangle + |58\rangle + 2 |77\rangle
+ 2 |80\rangle,
\nonumber\\
&&|\theta_{18}\rangle = - 2 |5\rangle - 2 |15\rangle + 2 |10\rangle 
+ 2 |11\rangle - 8 |12\rangle - |43\rangle - 8 |47\rangle - 8 |48\rangle 
- 2 |49\rangle + |50\rangle 
\nonumber\\
&&\hspace*{1cm}- |52\rangle + 4 |54\rangle + 2 |58\rangle +
4 |77\rangle + 12 |80\rangle + 8 |81\rangle - 2 |82\rangle + 2 |84\rangle,
\label{b1}
\end{eqnarray}

For examplification we present also two $U$ independent eigenfunctions
corresponding to non-zero eigenvalues, namely $+8$ (the vector $|v_1\rangle$),
and $-8$ (the vector $|v_2\rangle$).
\begin{eqnarray}
&&|v_1\rangle = |5\rangle - |15\rangle + |23\rangle + |26\rangle + |27 \rangle
+ |37\rangle + |38\rangle + |43\rangle + 2|44\rangle - |46\rangle - 2|47\rangle
+ |48\rangle 
\nonumber\\
&&\hspace*{1cm}- |49\rangle - |50\rangle - |51\rangle
- |52\rangle - 2 |53\rangle
- |54\rangle + 2 |67\rangle + 2 |68\rangle + 2 |69\rangle + |71\rangle 
- |72\rangle 
\nonumber\\
&&\hspace*{1cm}+ |73\rangle + 4 |77\rangle - 2 |78\rangle - 4 |80\rangle 
+ 2 |81\rangle + |82\rangle + |84\rangle - 2 |85\rangle,
\nonumber\\
&&|v_2\rangle = |5\rangle - |15\rangle - |23\rangle - |26\rangle - |27 \rangle
+ |37\rangle + |38\rangle + |43\rangle + 2|44\rangle - |46\rangle - 2|47\rangle
+ |48\rangle 
\nonumber\\
&&\hspace*{1cm}- |49\rangle - |50\rangle - |51\rangle 
- |52\rangle - 2 |53\rangle
- |54\rangle - 2 |67\rangle - 2 |68\rangle - 2 |69\rangle - |71\rangle 
+ |72\rangle 
\nonumber\\
&&\hspace*{1cm}- |73\rangle + 4 |77\rangle - 2 |78\rangle - 4 |80\rangle 
+ 2 |81\rangle + |82\rangle + |84\rangle - 2 |85\rangle,
\label{b2}
\end{eqnarray}


\newpage

\begin{figure}[h]
\centerline{\epsfbox{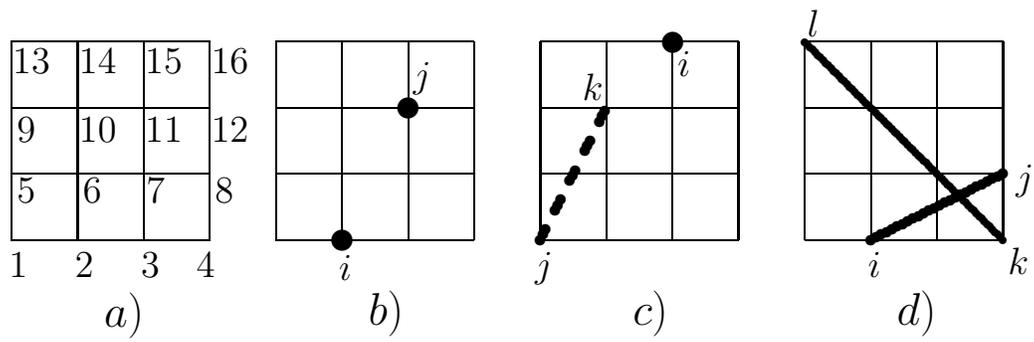}}
\caption{The site numbering for a $4\times 4$ cluster, and the possible
particle displacements for four electrons.}
\label{fig1}
\end{figure}

\newpage

\begin{figure}[h]
\centerline{\epsfbox{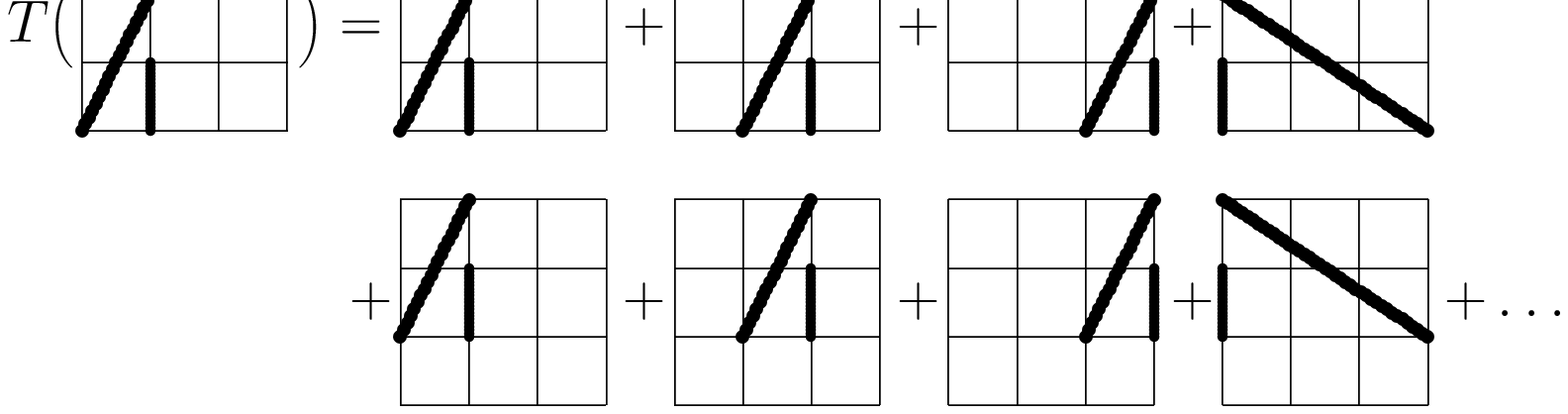}}
\caption{The effect of the translation operator $T$. The result in the right
hand side contains 16 terms, from which, the first 8 are ploted in the 
figure.}
\label{fig2}
\end{figure}

\newpage

\begin{figure}[h]
\centerline{\epsfbox{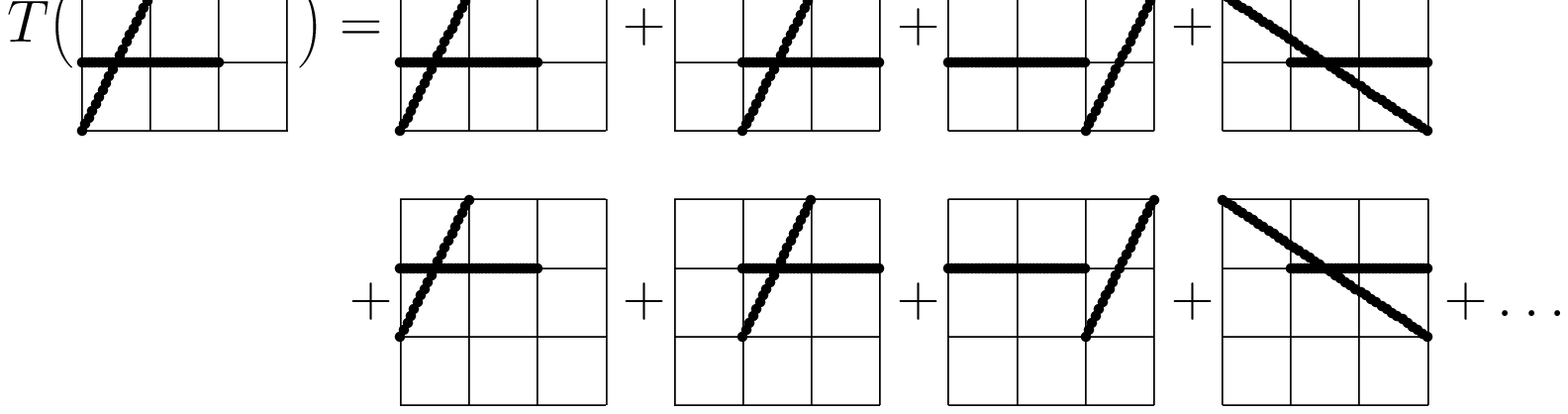}}
\caption{The effect of the translation operator $T$ leading to sign changes
in the mathematical expressions.}
\label{fig3}
\end{figure}
\newpage

\begin{figure}[h]
\centerline{\epsfbox{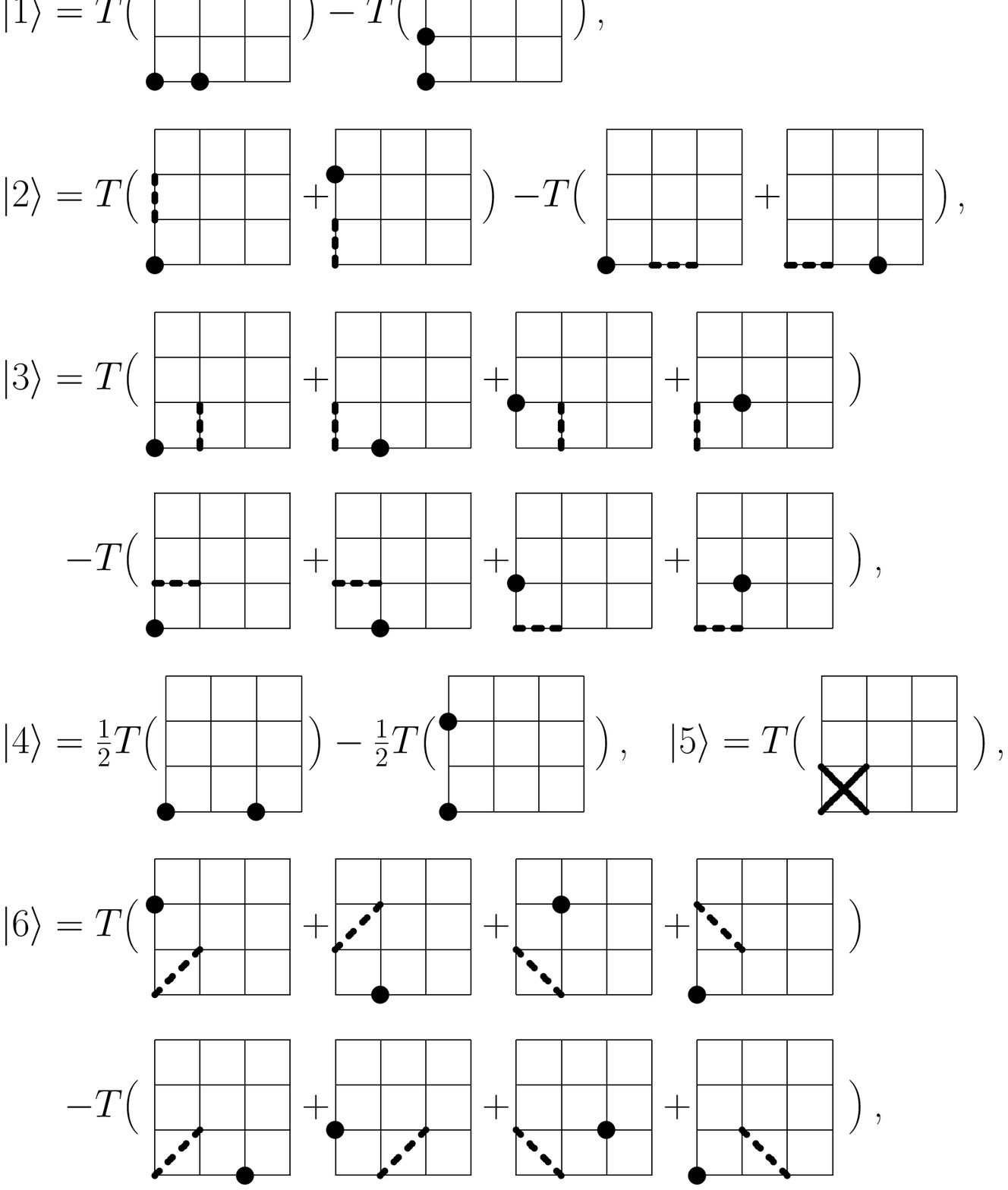}}
\caption{Base wave vectors $|1\rangle - |6\rangle$.}
\label{fig4}
\end{figure}

\newpage

\begin{figure}[h]
\centerline{\epsfbox{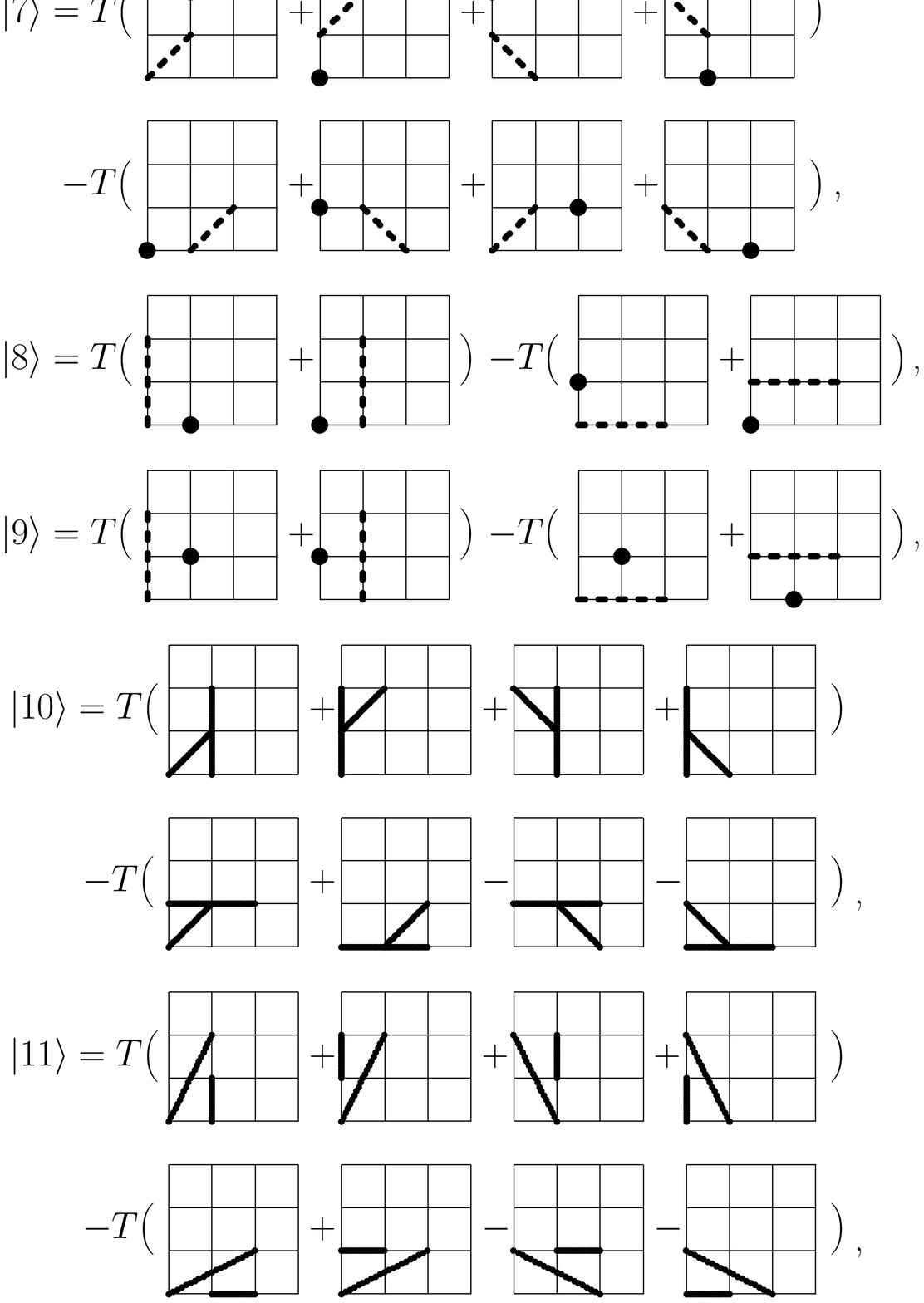}}
\caption{Base wave vectors $|7\rangle - |11\rangle$.}
\label{fig5}
\end{figure}

\newpage

\begin{figure}[h]
\centerline{\epsfbox{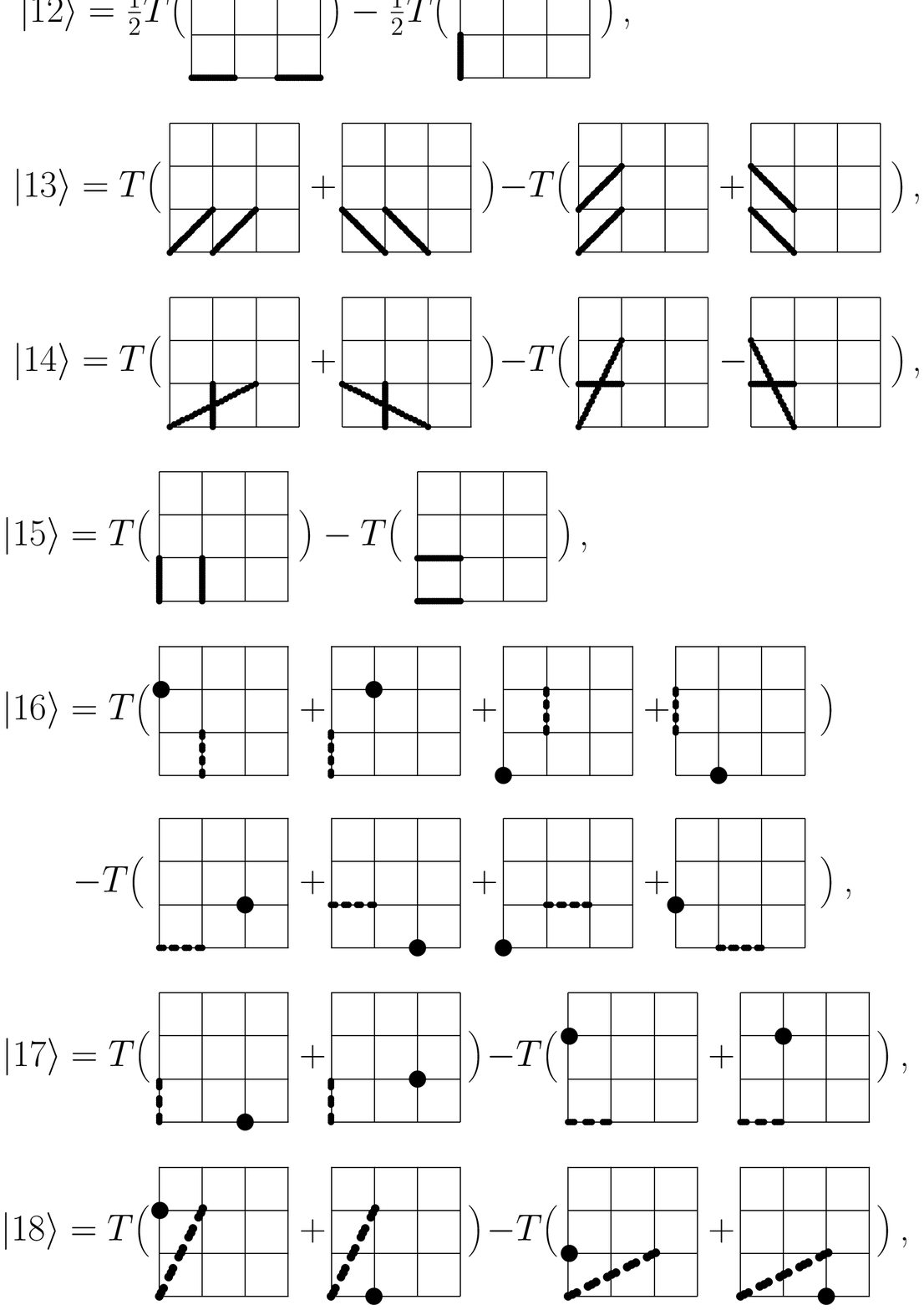}}
\caption{Base wave vectors $|12\rangle - |18\rangle$. }
\label{fig6}
\end{figure}

\newpage

\begin{figure}[h]
\centerline{\epsfbox{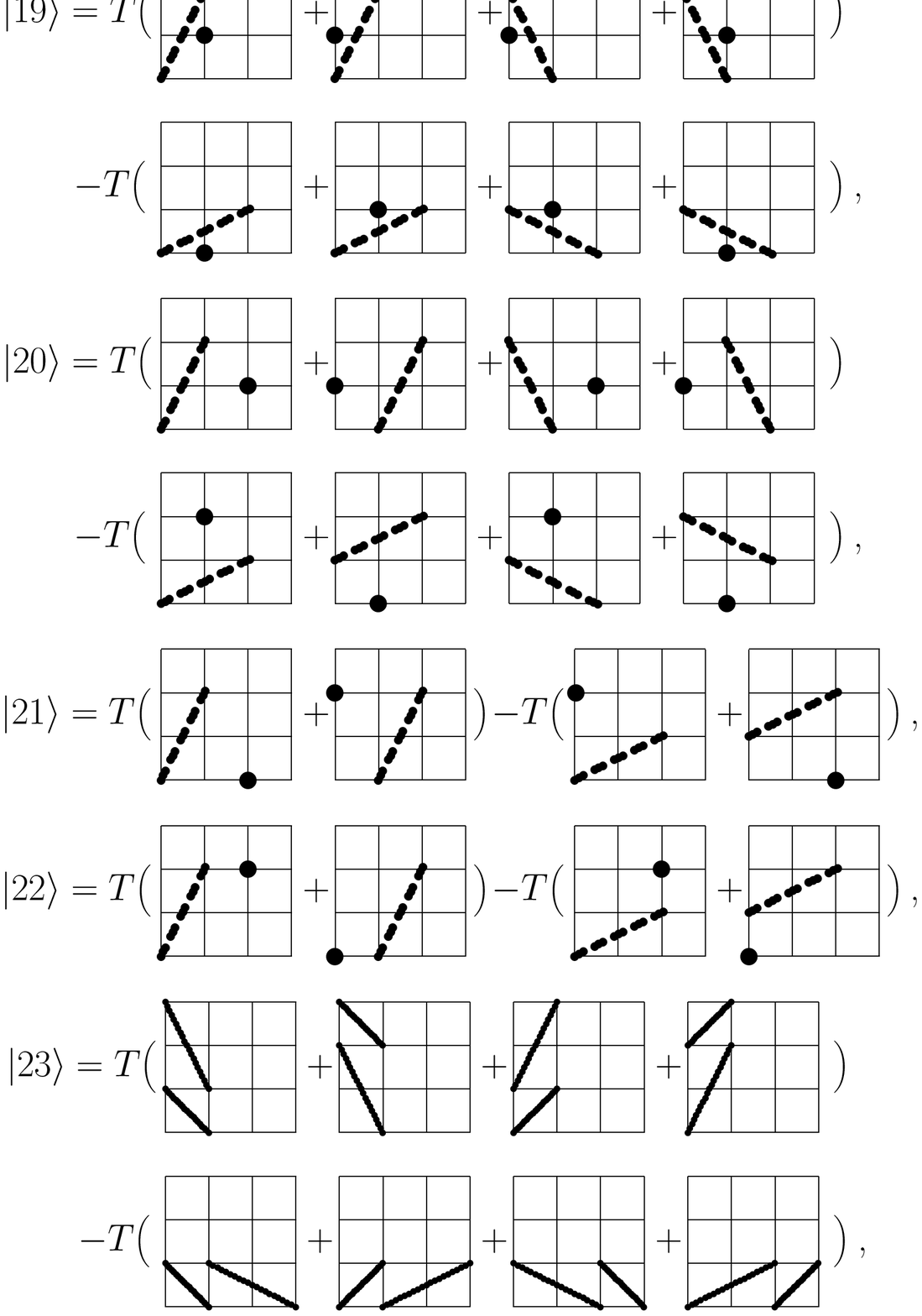}}
\caption{Base wave vectors $|19\rangle - |23\rangle$. }
\label{fig7}
\end{figure}

\newpage

\begin{figure}[h]
\centerline{\epsfbox{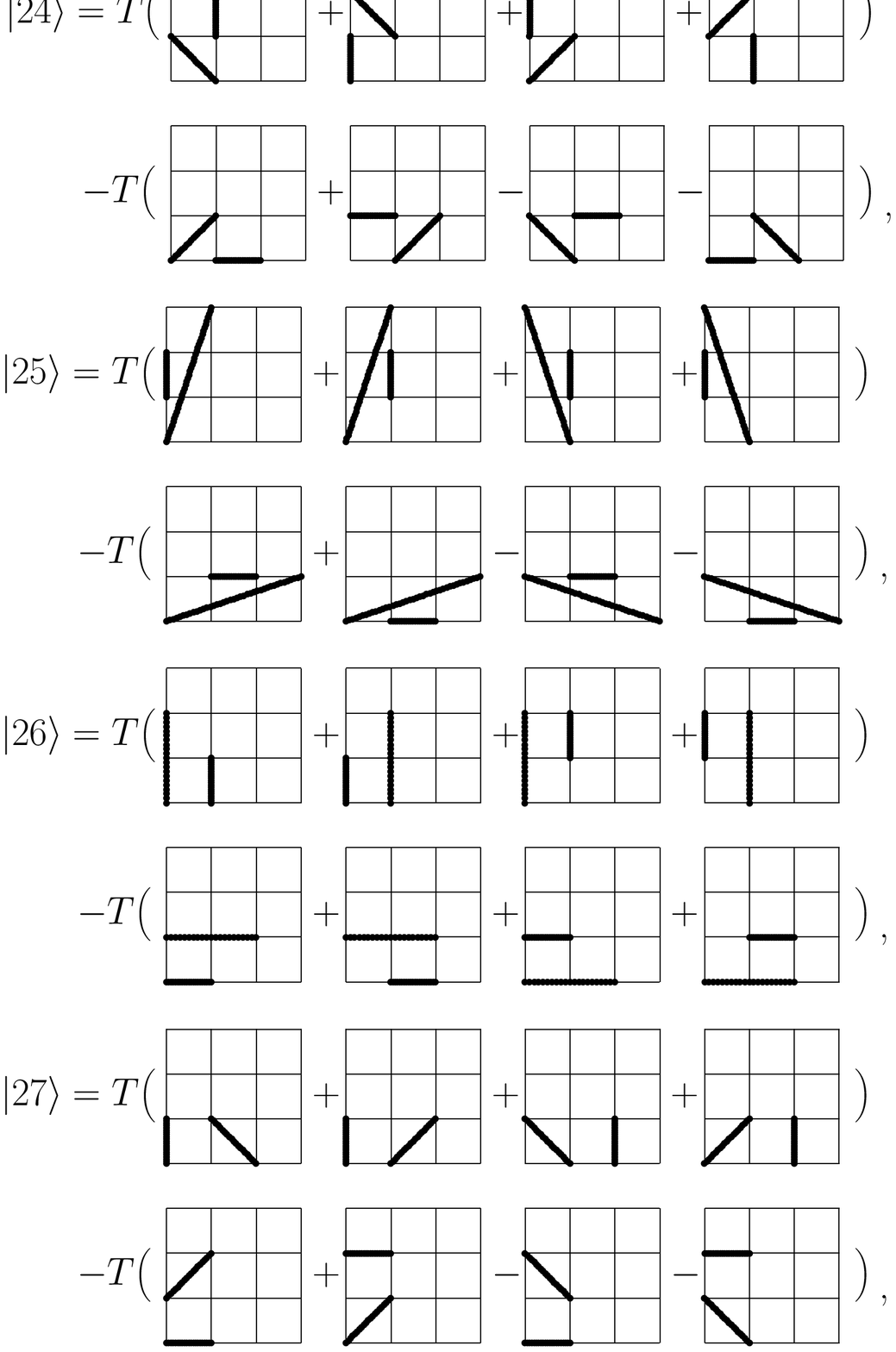}}
\caption{Base wave vectors $|24\rangle - |27\rangle$. }
\label{fig8}
\end{figure}

\newpage

\begin{figure}[h]
\centerline{\epsfbox{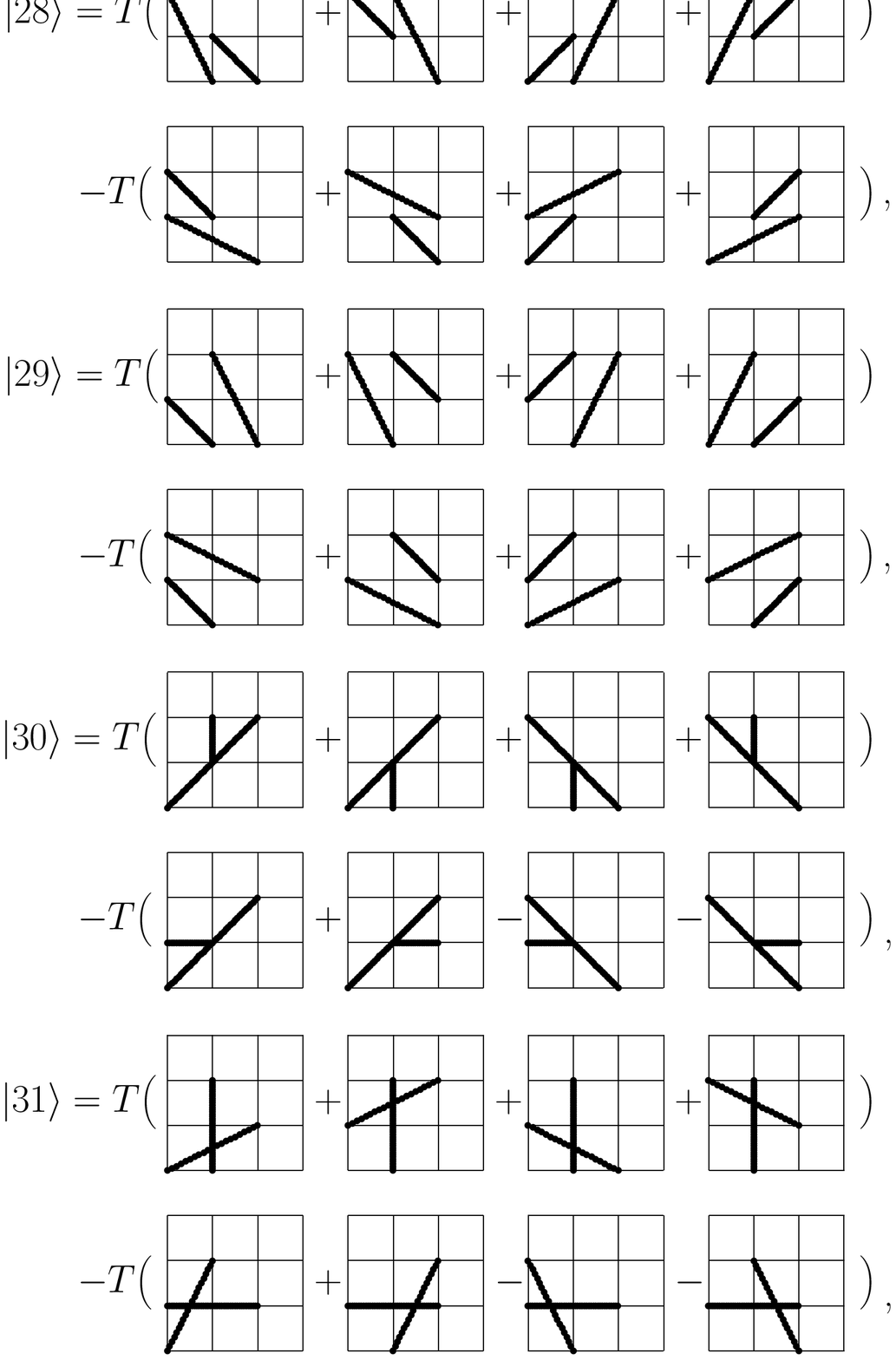}}
\caption{Base wave vectors $|28\rangle - |31\rangle$. }
\label{fig9}
\end{figure}

\newpage

\begin{figure}[h]
\centerline{\epsfbox{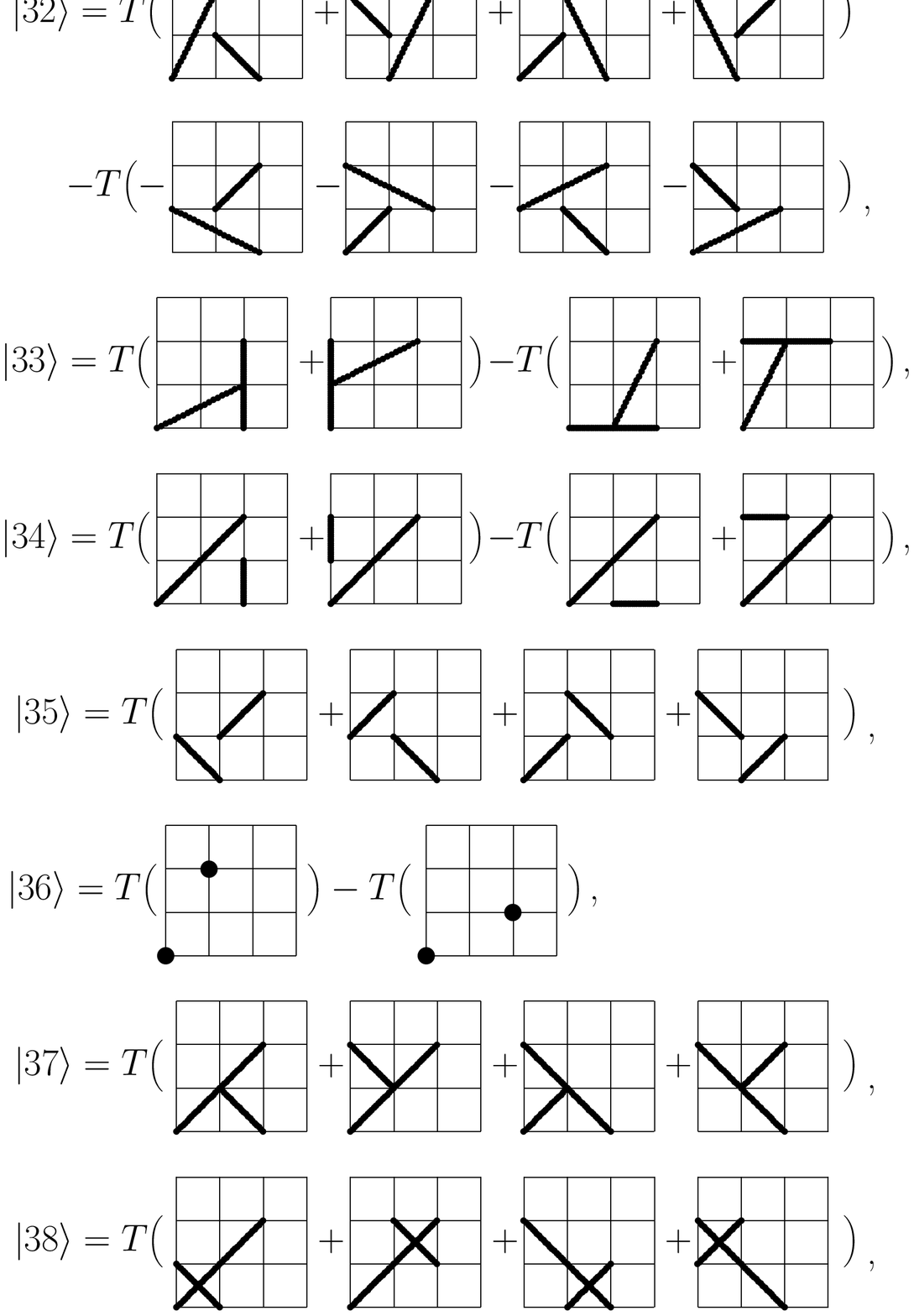}}
\caption{Base wave vectors $|32\rangle - |38\rangle$. }
\label{fig10}
\end{figure}

\newpage

\begin{figure}[h]
\centerline{\epsfbox{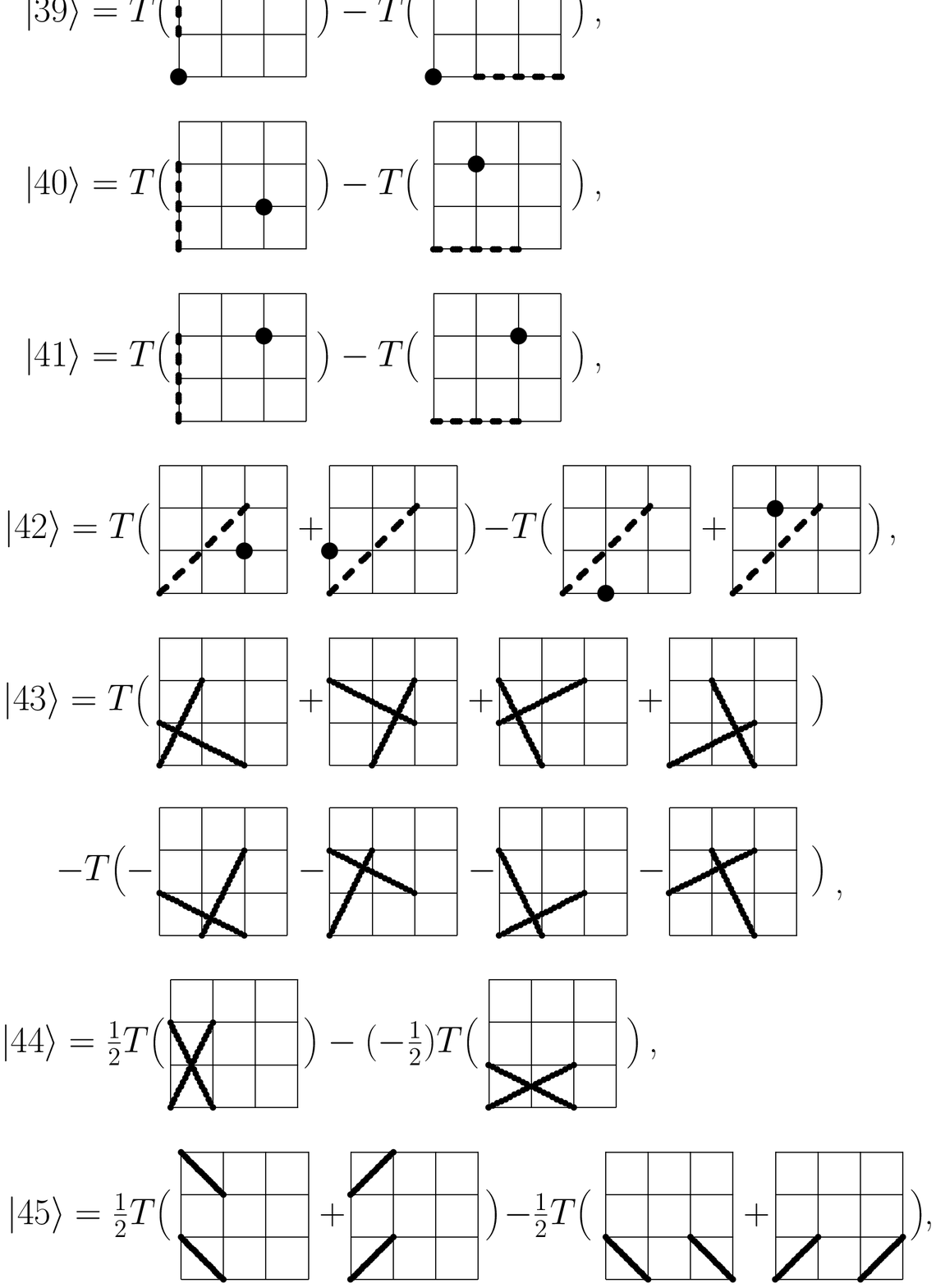}}
\caption{Base wave vectors $|39\rangle - |45\rangle$. }
\label{fig11}
\end{figure}

\newpage

\begin{figure}[h]
\centerline{\epsfbox{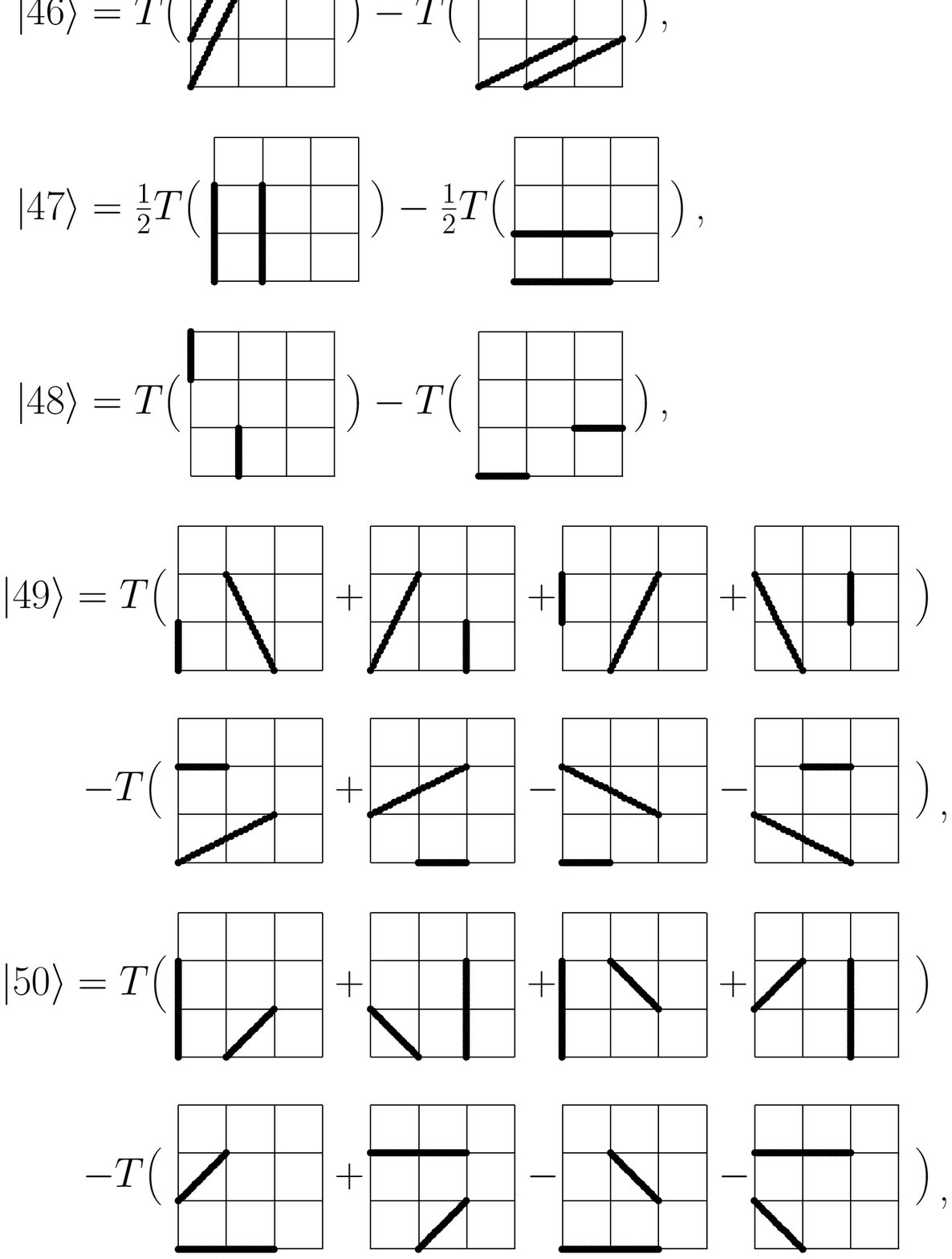}}
\caption{Base wave vectors $|46\rangle - |50\rangle$. }
\label{fig12}
\end{figure}

\newpage

\begin{figure}[h]
\centerline{\epsfbox{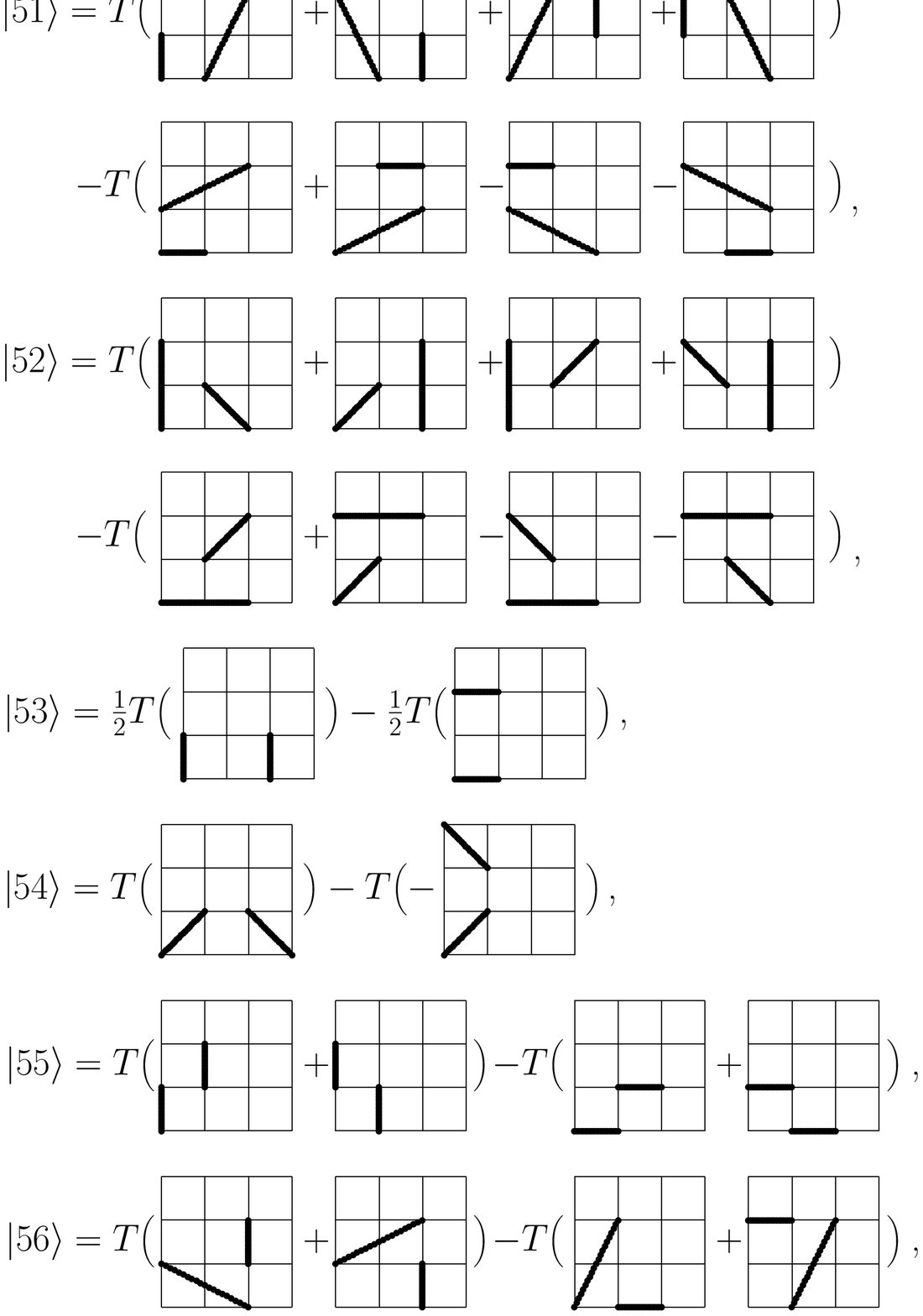}}
\caption{Base wave vectors $|51\rangle - |56\rangle$. }
\label{fig13}
\end{figure}

\newpage

\begin{figure}[h]
\centerline{\epsfbox{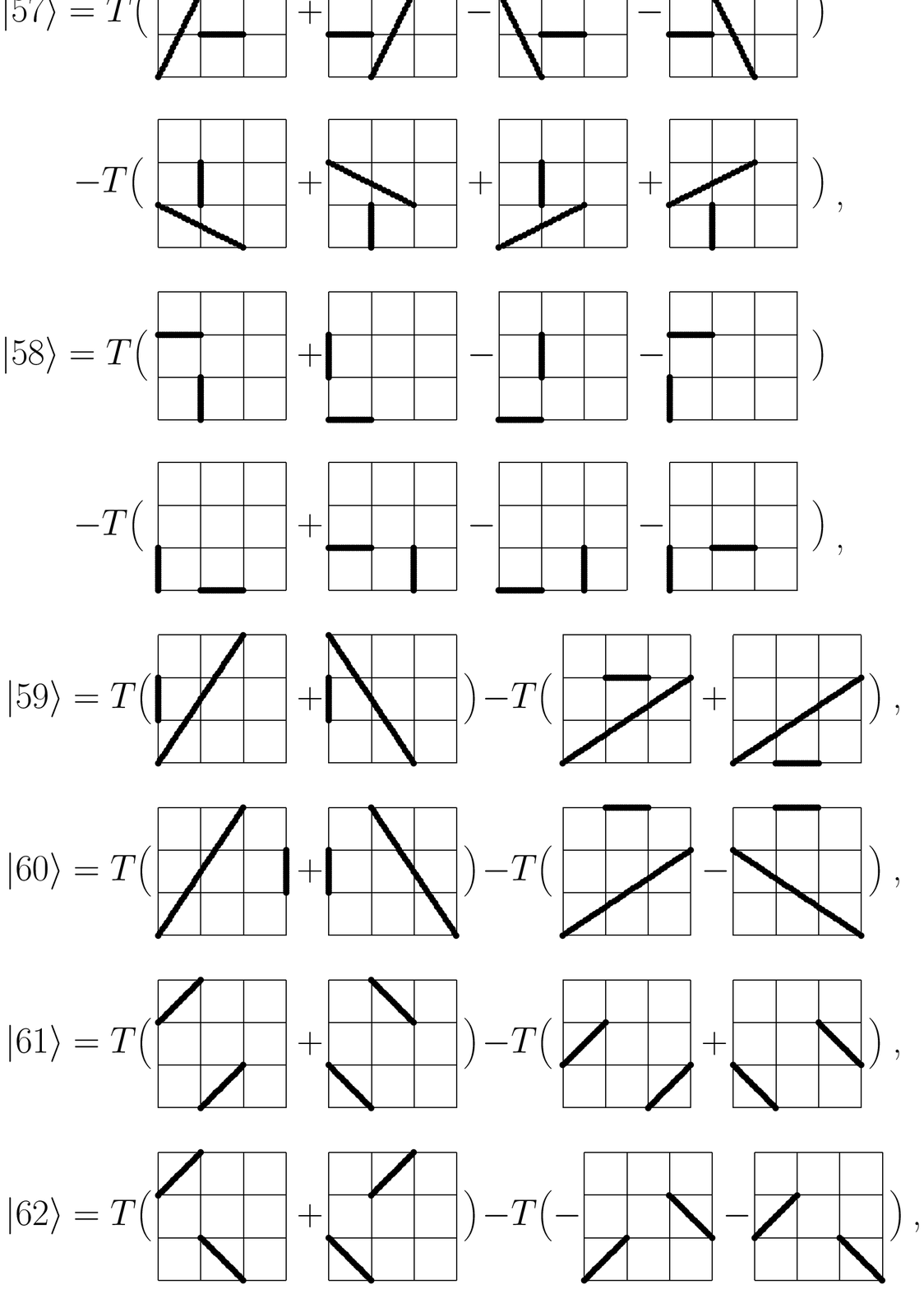}}
\caption{Base wave vectors $|57\rangle - |62\rangle$. }
\label{fig14}
\end{figure}

\newpage

\begin{figure}[h]
\centerline{\epsfbox{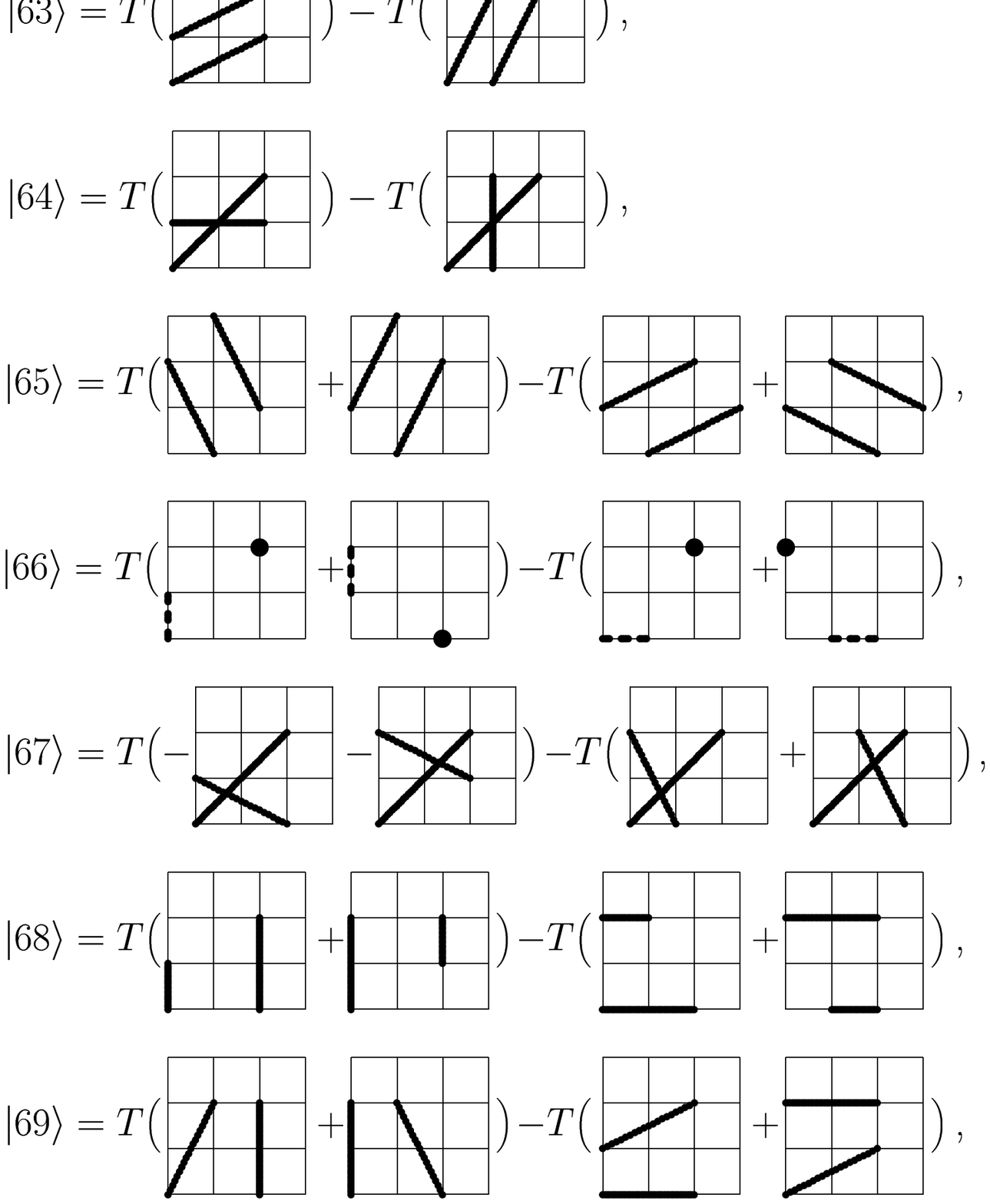}}
\caption{Base wave vectors $|63\rangle - |69\rangle$. }
\label{fig15}
\end{figure}

\newpage

\begin{figure}[h]
\centerline{\epsfbox{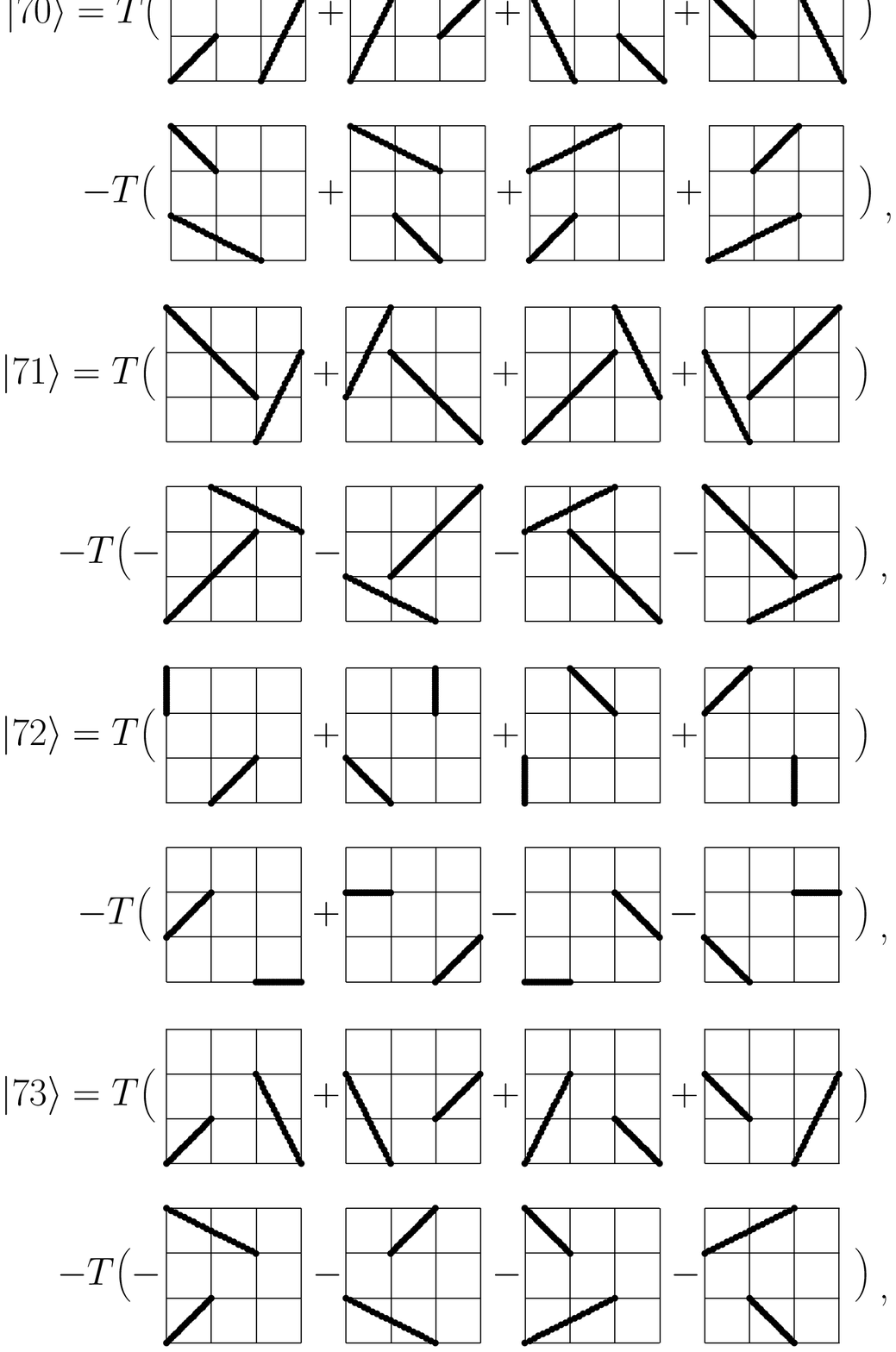}}
\caption{Base wave vectors $|70\rangle - |73\rangle$. }
\label{fig16}
\end{figure}

\newpage

\begin{figure}[h]
\centerline{\epsfbox{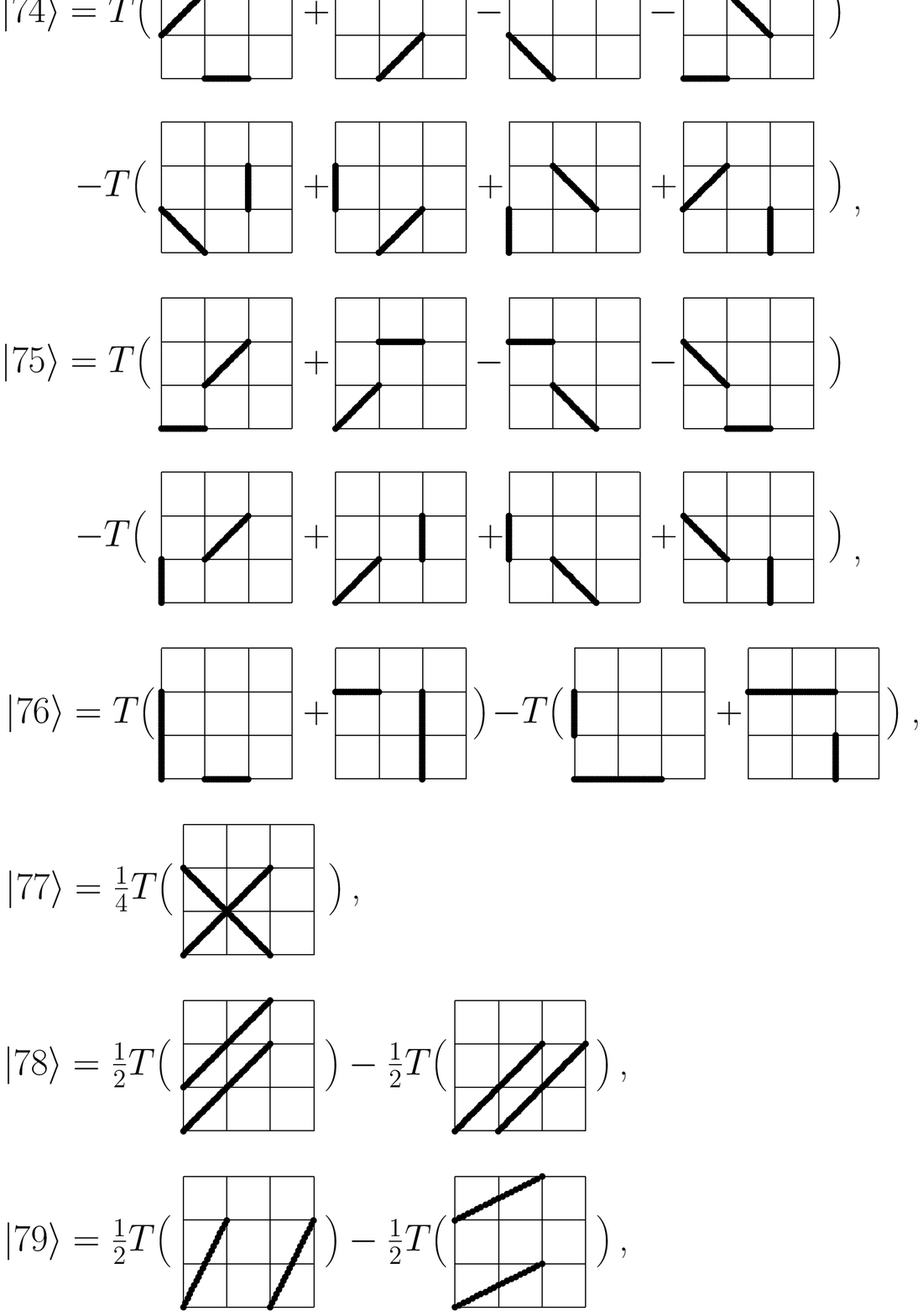}}
\caption{Base wave vectors $|74\rangle - |79\rangle$. }
\label{fig17}
\end{figure}

\newpage

\begin{figure}[h]
\centerline{\epsfbox{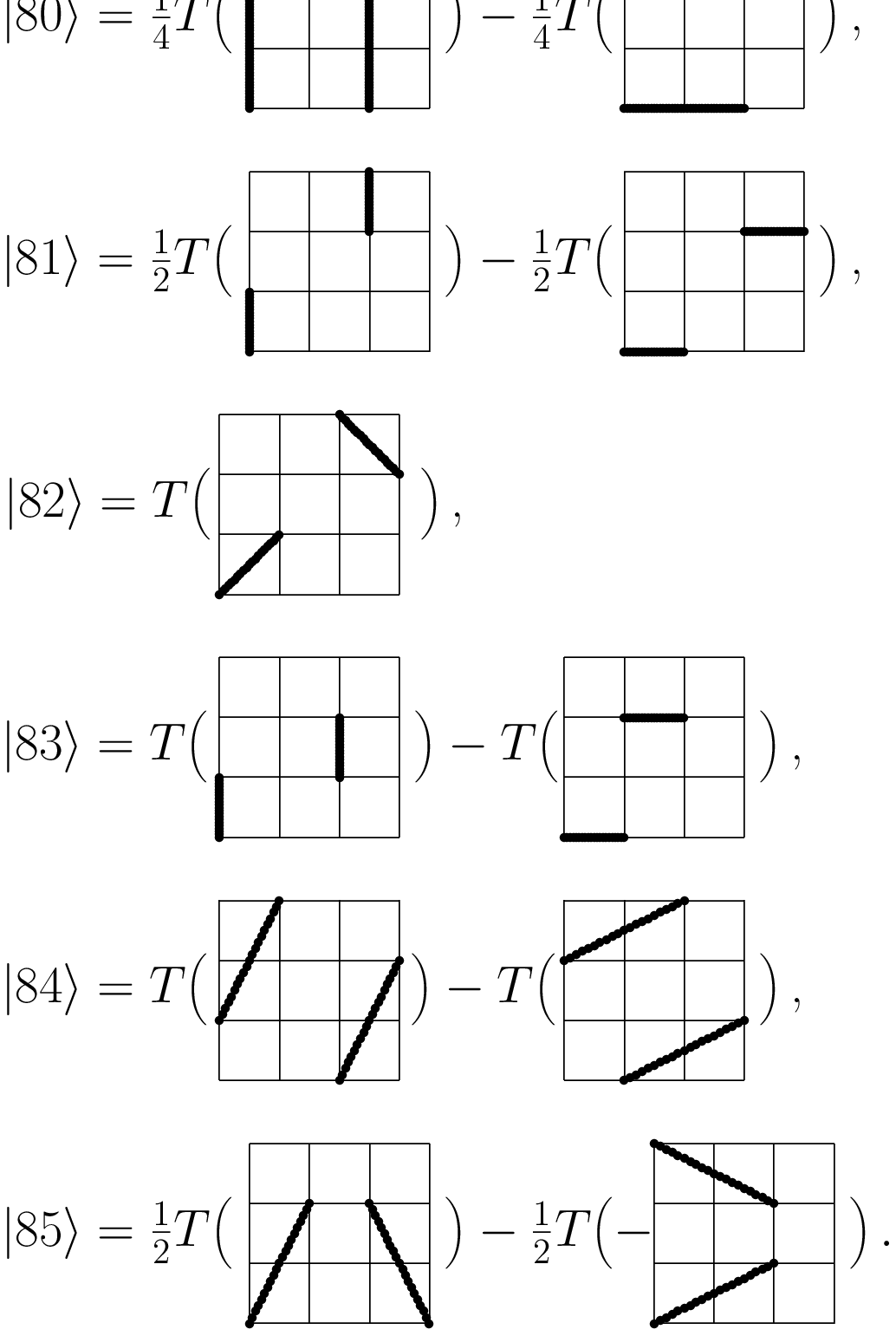}}
\caption{Base wave vectors $|80\rangle - |85\rangle$. }
\label{fig18}
\end{figure}

\newpage


\begin{thebibliography}{dummy}

\bibitem[Acquarone et al.(1999)]{E2D17}ACQUARONE, M., CUCCO, M., NOCE, C., and
ROMANO, A., 1999, Physica {\bf B261}, 725.

\bibitem[Anderson(1995)]{s2}ANDERSON, P., W., 1995, Science {\bf 268}, 1154.

\bibitem[Baeriswyl et al.(1995)]{E2D3}BAERISWYL, D., CAMPBELL, D., K.,
CARMELO, J., M., P., GUINEA, F., and LOUIS, E., Eds., 1995,
{\em The Hubbard Model, Its Physics and Mathematical Physics}, NATO ASI 
Series B, vol. 343, Plenum Press, New York and London.

\bibitem[Balzarotti et al.(2004)]{c4}BALZAROTTI, A., et. al., 2004,
cond-mat/0411101.

\bibitem[Boyaci, Gedik and Kulik(2000)]{End3}BOYACI, H., GEDIK, Z., and
KULIK, I., O., 2000, Jour. of Supercond. {\bf 13}, 1031; ibid. 2001,
{\bf 14}, 133.

\bibitem[Bruus and d'Auriac(1996)]{E2D14}BRUUS, H., and D'AURIAC, J., C., A.,
1996, Europhys. Lett. {\bf 35}, 321.

\bibitem[Bruus and d'Auriac(1997)]{E2D15}BRUUS, H., and D'AURIAC, J., C., A.,
1997, Phys. Rev. {\bf B55}, 9142.

\bibitem[Busser, Moreno and Dagotto(2004)]{End2}BUSSER, C., A., MORENO, A.,
and DAGOTTO, E., 2004, Phys. Rev. {\bf B70}, 035402.

\bibitem[Chen and Mei(1980)]{E2D7}CHEN, L., and MEI, C., 1989,
Phys. Rev. {\bf B39}, 9006.

\bibitem[Chiappe et al.(1999)]{End8}CHIAPPE, G., BUSSER, C., ANDA, E., V.,
and FERRARI, V., 1999, Jour. of Phys. {\bf C11}, 5237.

\bibitem[Cini, Perfetto and Stefanucci(2001)]{c1}CINI, M., PERFETTO, E., and
STEFANUCCI, G., 2001, Eur. Phys. Jour. {\bf B20}, 91.

\bibitem[Cini and Stefanucci(2001)]{c3}CINI, M., and STEFANUCCI, G., 2001,
Jour. of Phys. {\bf C13}, 1279.

\bibitem[Eckl, Hanke and Arrigoni(2003)]{s5}ECKL, T., HANKE, W., and
ARRIGONI, E., 2003, Phys. Rev. {\bf B68}, 014505.

\bibitem[Eroles, Batista and Aligia(1999)]{End7}EROLES, J., BATISTA, C., D.,
and ALIGIA, A., A., 1999, Phys. Rev. {\bf B59}, 14092.

\bibitem[Fabrizio, Parola and Tosatti(1991)]{E2D6}FABRIZIO, M., PAROLA, A., and
TOSATTI, E., 1991, Phys. Rev. {\bf B44}, 1033.

\bibitem[Falicov and Proetto(1993)]{E2D5}FALICOV, L., M., and PROETTO, C., R.,
1993, Phys. Rev. {\bf B47}, 14407.

\bibitem[Fano, Ortolani and Parola(1992)]{E2D13}FANO, G., ORTOLANI, F., and
PAROLA, A., 1992, Phys. Rev. {\bf B46}, 1048.

\bibitem[Feng(2003)]{s3}FENG, S., 2003, Phys. Rev. {\bf B68}, 184501.

\bibitem[Galan and Verges(1991)]{E2D12}GALAN, J., and VERGES, J., A., 1991,
Phys. Rev. {\bf B44}, 10093.

\bibitem[Halfpap(2001)]{d5}HALFPAP, O., 2001, Annalen der Physik {\bf 10}, 
623.

\bibitem[Helberg(2001)]{End6}HELBERG, C., S., 2001, Jour. Appl. Phys. 
{\bf 89}, 6627.

\bibitem[Hirsch(2004)]{s1}HIRSCH, J., E., 2000, Physica {\bf C341-348}, 213;
and ibid 1992, {\bf C199}, 305.

\bibitem[Hubbard(1963)]{E2D1}HUBBARD, J., 1963, Proc. R. Soc. London
{\bf A276}, 238.

\bibitem[Kochereshko et al.(2003)]{i2}KOCHERESHKO, V., P., et al., 2003,
Physica {\bf E17}, 197.

\bibitem[Lieb(1994)]{E2D4}LIEB, E., H., 1994,
in {\em Proceedings of the XIth International Congress of Mathematical 
Physics}, Paris, edited by D. Iagolnitzer, International Press, Paris, 
pp. 392.

\bibitem[Lieb and Wu(1968)]{E2D2}LIEB, E., H., and WU, F., Y., 1968,
Phys. Rev. Lett. {\bf 20}, 1445.

\bibitem[Louis et al.(1992)]{E2D9}LOUIS, E., GALAN, J., GUINEA, F., VERGES,
J., A., and FERRER, J., 1992, Phys. Stat. Solidi {\bf B173}, 715.

\bibitem[Maksym et al.(2000)]{i1}MAKSYM., P., A., IMAMURA, H., MALLON, G., P.,
and AOKI, H., 2000, Jour. of Phys. {\bf C12}, R299.

\bibitem[van der Marel et al.(2003)]{s6}MAREL, VAN, DER, D., MOLEGRAAF, H., J.,
A., PRESURA, C., and SANTOSO, I., 2003, cond-mat/0302169.

\bibitem[Mattis(1993)]{End9}MATTIS, D., C., 1993, {\em An Encyclopedia of 
Exactly Solved Models in One Dimension: The Many-Body Problem}, World 
Scientific, London.

\bibitem[Mei and Chen(1988)]{E2D7a}MEI, C., and CHEN, L., 1988,
Zeit.Phys. {\bf B72}, 429.

\bibitem[Metzner and Vollhardt(1989)]{vol}METZNER, W., and VOLLHARDT, D., 1989,
Phys. Rev. {\bf B39}, 4462.

\bibitem[Okada (2004)]{End4}OKADA, K., 2004, Jour. Phys. Soc. Jpn. 
{\bf 73}, 1681.

\bibitem[Papavassiliou and Yartsev(1992)]{E2D5a}PAPAVASSILIOU, G., C., and
YARTSEV, V., M., 1992, Chem. Phys. Lett., {\bf 200}, 209.

\bibitem[Parola et al.(1990)]{E2D8}PAROLA, A., SORELLA, S., PARRINELLO, M.,
and TOSATTI, E., 1990, in {\em Dynamics of Magnetic 
Fluctuations in High Temperature Superconductors}, edited by G. Reiner, 
P. Horsch, and G. Psaltakis, Plenum, New York.

\bibitem[Parola et al.(1991)]{E2D11}PAROLA, A., SORELLA, S., PARRINELLO, M.,
and TOSATTI, E., 1991, Phys. Rev. {\bf B43}, 6190.

\bibitem[Perfetto and Cini(2004)]{c2}PERFETTO, E., and CINI, M., 2004,
Jour. of Phys. {\bf C16}, 4845.

\bibitem[Sackett et al.(2000)]{i5}SACKETT, C., A., 2000, 
Nature {\bf 404}, 256.

\bibitem[Saito et al.(2001)]{End1}SAITO, G., HIRATE, S., NISHIMURA, K., and
YAMOCHI, H., 2001, Jour. of Mater. Chem. {\bf 11}, 723.

\bibitem[Sorella and Hubsh(2000)]{obs}SORELLA, S., and HUBSH, A., 2000,
{\em We would like to thank to S. Sorella and A. Hubsh for providing us 
numerical exact diagonalization results for the test of the deduced system of 
equations}.

\bibitem[Srinitiwarawong and Gehring(2002)]{End5}SRINITIWARAWONG, C., and
GEHRING, G., A., 2002, Jour. of Phys. {\bf C14}, 11589.

\bibitem[Tjemberg(1998)]{E2D16}TJEMBERG, O., 1998, Jour. Math. Phys. 
{\bf 39}, 6416.

\bibitem[Yokoyama et al.(2004)]{s4}YOKOYAMA, H., TANAKA, Y., OGATA, M., and
TSUCHIURA, H., 2004, Jour. Phys. Soc. Jpn. {\bf 73}, 1119.

\bibitem[Zhang and Henley(2004)]{E2D18}ZHANG, N., G., and HENLEY, C., L.,
2004, Eur. Phys. Jour. {\bf B38}, 409.

\bibitem[Zhu and Jiang(1993)]{E2D10}ZHU, H., Y., and JIANG, Y., S., 1993,
Acta Chim. Sinca {\bf 51}, 527.

\end{thebibliography}
\end{document}